\documentclass{tLCT2e}

\begin{document}

\articletype{INVITED ARTICLE}

\title{Shape-dynamic growth, structure, and elasticity of homogeneously oriented spherulites in an isotropic/smectic-A mesophase transition}

\author{Nasser Mohieddin Abukhdeir$^\ast$ and Alejandro D. Rey \thanks{$^\ast$Corresponding author. Email: nasser.abukhdeir@mcgill.ca}\\{\vbox{\vspace{12pt} {\em{Department of Chemical Engineering, McGill University, 3610 University St., Montreal, Quebec, Canada}}\\\vspace{6pt}\received{\date}}} }

\maketitle

\begin{abstract}
A Landau-de Gennes model that integrates the nematic quadrupolar tensor order parameter and complex smectic-A order parameters is used to simulate the two-dimensional growth of an initially homogeneous smectic-A spherulite in an isotropic matrix.  These simulations are performed in the shape-dynamic (nano-scale) regime of growth under two material conditions: isotropic nematic elasticity and equal splay-bend nematic elasticity.  A comparison of the growth kinetics, spherulite morphology, interfacial/bulk energy landscapes between both cases is made showing that equal nematic splay-bend elasticity is required to reproduce past experimental and theoretical observations.  Additionally, a previously unknown undulation instability during spherulite growth is found which, in conjunction with preferred planar anchoring and defect shedding mechanisms at micron length scales, could explain the formation mechanism of focal conic curvature defects and ultimately smectic-A ``batonnet'' structures observed experimentally.

\bigskip
\end{abstract}

\section{Introduction}

The study of liquid crystalline materials has had a rapid and profound technological impact on civilisation in the past century.  In the overarching field of Soft Matter, the study of liquid crystals has contributed fundamental advances in technological areas such as display technology and high-performance materials.  Additionally, interest is quickly emerging in an area in which the understanding of liquid crystals is being recognised as crucial: biological systems \cite{Lockwood2008}, and Nature as a whole.  Self-organisation is a basis on which all biological systems have developed and evolved and the presence of liquid crystal phases in the very building blocks of Nature, the cell's phospholipid bilayer, is self-evident \cite{Fisch2004,Lockwood2008}.  As our understanding of biological systems increases, more and more instances of liquid crystal phases are found \cite{Rizvi2003}: in muscle tissue \cite{Aldoroty1987}, the development of bone \cite{Fisch2004}, and even in the beginnings of life itself \cite{Nakata2007}.

Over the course of almost four decades, the vast contributions of P.G. de Gennes to the area of liquid crystals have enabled the advancement of knowledge up to this point.  One of these contributions was his theoretical work on the smectic-A mesophase.  His discovery, simultaneously with McMillan,  of the analogy between superconductors and smectic-A phase ordering has opened up an avenue for study of this mesophase via modelling and simulation.  The key role that this mesophase plays in biological systems alone shows the importance of this single contribution of de Gennes.  In his Nobel lecture in 1991, he states that ``smectics bring me naturally to another important feature of complex fluids - namely that, in our days, it is sometimes possible to create new forms of matter'' which alludes to the types of applications of smectics that could be developed in the future and to their roles in biological systems.

The most simple of the smectics is the smectic-A mesophase, which displays lamellar translational ordering, in addition to the orientational ordering of nematics.  Recently an increasing amount of interest in this mesophase, in particular of materials exhibiting a direct isotropic/smectic-A (disordered/ordered) transition, has resulted in many experimental and theoretical results.  Nonetheless, the understanding of this mesophase is in a nascent stage.  Much of this is due to the time and length scales at which the structures and dynamics occur being on the nano-scale.  These properties provide a obvious application of theoretical study through modelling and simulation in order to both enhance experimental research and make predictions independently.

A fascinating range of liquid crystal growth morphologies have been the focus on much study \cite{Rey2007}.  The smectic-A mesophase, with its lamellar ordering on the molecular scale, exhibits growth, defect, and texture phenomena not seen in the nematic phases, including the lamellar-like cholesteric mesophase.  Focusing on the growth phenomena solely, on transition from the isotropic/disordered phase, a variety of self-assembled smectic-A structures have been observed to form.  These unique morphologies can be attributed to complex dynamics involving interfacial tension anisotropy, which results in preferred anchoring, and bulk texturing.

The current study is a part of an overall effort to understand kinetics, dynamics, and morphology of the direct isotropic/smectic-A liquid crystalline phase transition.  The theoretical focal point of this research is a high-order Landau-de Gennes type phenomenological model of Mukherjee, Pleiner, and Brand \cite{deGennes1995,Mukherjee2001}.  This high-order model incorporates much of the key physics involved in the direct isotropic/smectic-A transition which occur on multiple scales and involve multiple types of phase-ordering: orientational and translational.  In this effort a comprehensive approach has been developed for the determination of phenomenological parameters for the model and efficient phase diagram computation \cite{Abukhdeir2007}.  Following this, numerical simulation was used to study phase transition kinetics and defect dynamics \cite{Abukhdeir2008a}, surface effects \cite{Abukhdeir2008}, and most recently spherulite growth \cite{Abukhdeir2008c}.  In addition to these simulation studies, theoretical contributions have been made to the study of smectic-A filamentary growth and buckling \cite{Rey2008,Rey2008a}.

Past work on smectic-A spherulite growth has focused on initially radially oriented nuclei \cite{Abukhdeir2008c}.  The current work studies an alternate initial configuration where the nucleus is homogeneously, or ``ideally,'' oriented.  This and past work \cite{Abukhdeir2008,Abukhdeir2008a,Abukhdeir2008c,Rey2008a} focuses on rod-like low molecular mass liquid crystals and phenomenological parameters are based, in part, on experimental data from 12CB (dodecyl-cyanobiphenyl).  This work neglects nucleation mechanisms, thermal fluctuations, heat of transition, impurities, and convective flow while taking into account energetically the inter-coupling between orientational/translational order and variation of smectic layer spacing.  The objectives of this simulation study are:
\begin{enumerate}
\item Determine the dynamic growth morphology of an initially homogeneous smectic spherulite with no preferred anchoring at the isotropic/smectic-A interface (isotropic nematic elasticity).
\item Determine the dynamic growth morphology of an initially homogeneous smectic spherulite with preferred planar anchoring (as is observed experimentally) at the isotropic/smectic-A interface (equal bend-splay nematic elasticity).
\item Compare these simulation predictions with simplified shape equation models.
\end{enumerate}
The paper begins with brief background on liquid crystals and phase-ordering transitions.  The model is then presented and details of the simulation method/ conditions detailed.  A brief discussion of the connection between two- and three-dimensional simulation is then given using a less complex example of initially homogeneous spherulite growth in the isotropic/nematic transition.  Finally, results of the work are presented and conclusions made.

\section{Background}

\subsection{The first-order isotropic/smectic-A transition}

\begin{figure}
\begin{center}
\includegraphics[width=10cm]{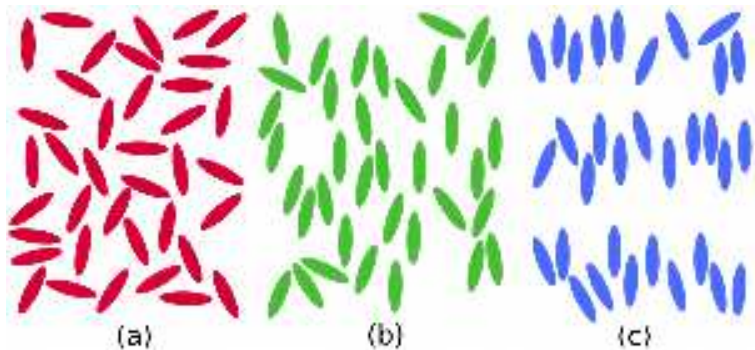}
\caption{schematics of the (a) isotropic, (b) nematic, and (c) smectic-A phases. \label{fig:lcorder}}
\end{center}
\end{figure}

This work focuses the study of rod-like thermotropic liquid crystals which exhibit a first-order isotropic/smectic-A mesophase transition.  An unordered liquid, where there is neither orientational nor translational order (apart from an average intermolecular separation distance) of the molecules, is referred to as isotropic.  Liquid crystalline order involves partial orientational order (nematics) and, additionally, partial translational order (smectics and columnar mesophases).  The simplest of the smectics is the smectic-A mesophase, which exhibits one-dimensional translational order in the direction of the preferred molecular orientational axis, which can be thought of as layers of two-dimensional fluids stacked upon each other.  Other more complex types of smectics exist, for example tilted smectic/smectic-C and hexatic smectic/smectic-B mesophases.  In this context, the relative simplicity of the smectic-A mesophase makes it the ideal starting point and, subsequently, a template phase for the vast set of self-assembled lamellar systems \cite{Harrison2000}.  Schematic representations of these different types of ordering are shown in figure \ref{fig:lcorder}.

\begin{figure}
\begin{center}
\includegraphics[width=10cm]{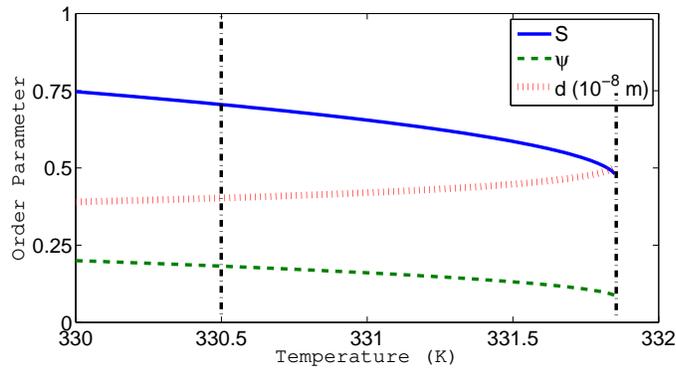}
\caption{Computed phase diagram for 12CB \cite{Abukhdeir2007} with nematic scalar order parameter (solid line), smectic scalar order parameter (dotted line), and the normalised wave vector (stippled line).  The isotropic/smectic-A coexistence region is enclosed by the vertical dashed lines, indicating the temperature range over which the isotropic/smectic-A phases coexist in stable/meta-stable states \cite{Coles1979a,Abukhdeir2007}. \label{fig:12CB}}
\end{center}
\end{figure}

Due to the first-order nature of the isotropic/smectic-A transition, a coexistence temperature interval exists where both the isotropic and smectic-A phase are either stable or metastable.  The phase diagram computation method for the model used in this work was developed previously \cite{Abukhdeir2007} and the resulting phase diagram for the phenomenological system used in this work is presented in figure \ref{fig:12CB}.

\section{Modelling and Simulation}

\subsection{The Landau-de Gennes model}

Theoretical characterisation of mesophase order is accomplished using order parameters that adequately capture the physics involved. These order parameters typically have an amplitude and phase associated with them. In order to characterise the partial orientational order of the nematic phase, a second order symmetric traceless tensor can be used \cite{deGennes1995}:
\begin{equation} \label{eqn:nem_order_param}
\bm{Q} = S \left(\bm{nn} - \frac{1}{3} \bm{I}\right) + \frac{1}{3} P \left( \bm{mm} - \bm{ll}\right)
\end{equation}                   
where $\mathbf{n}/\mathbf{m}/\mathbf{l}$ are the eigenvectors of $\bm{Q}$, which characterise the average molecular orientational axes, and $S/P$ are scalars which characterise the extent to which the molecules conform to the average orientational axes \cite{Rey2002,Yan2002,Rey2007}.  Uniaxial order is characterised by $S$ and $\bm{n}$, which correspond to the maximum eigenvalue (and its corresponding eigenvector) of $\bm{Q}$, $S= \frac{3}{2} \mu_n$.  Biaxial order is characterised by $P$ and $\bm{m}/\bm{l}$, which correspond to the lesser eigenvalues and eigenvectors, $P = -\frac{3}{2}\left(\mu_m - \mu_l\right)$.

The smectic-A mesophase has one-dimensional translational order in addition to the orientational order found in nematics.  Characterising this mesophase can be accomplished through the use of primary (orientational) and secondary (translational) order parameters together \cite{Toledano1987}.  A complex order parameter can be used to characterise translational order \cite{deGennes1995}:
\begin{equation} \label{eqsmec_order_param}
\Psi = \psi e^{i \phi}
\end{equation}
where $\phi$ is the phase, $\psi$ is the scalar amplitude of the density modulation.  The density wave vector, which describes the average orientation of the smectic-A density modulation, is defined as $\mathbf{a} = \nabla \phi / |{\nabla \phi}|$.  The smectic scalar order parameter $\psi$ characterises the magnitude of the density modulation, and is used in a dimensionless form in this work.  In the smectic-A mesophase the preferred orientation of the wave vector is parallel to the average molecular orientational axis, $\mathbf{n}$.

A Landau-de Gennes type model for the first order isotropic/smectic-A phase transition is used that was initially presented by Mukherjee, Pleiner, and Brand \cite{deGennes1995,Mukherjee2001} and later extended by adding nematic elastic terms \cite{Brand2001,Mukherjee2002a,Biscari2007}:
\begin{eqnarray} \label{eq:free_energy_heterogeneous}
f - f_0 &=&\frac{1}{2} a \left(\bm{Q} : \bm{Q}\right) - \frac{1}{3} b \left(\bm{Q}\cdot\bm{Q}\right) : \bm{Q} + \frac{1}{4} c \left(\bm{Q} : \bm{Q}\right)^2 + \frac{1}{2} \alpha \left|\Psi\right|^2 + \frac{1}{4} \beta \left|\Psi\right|^4 \nonumber\\
&&- \frac{1}{2} \delta \psi^2 \left(\bm{Q} : \bm{Q}\right) - \frac{1}{2} e \bm{Q}:\left(\bm{\nabla} \Psi\right)\left(\bm{\nabla} \Psi^*\right) \nonumber\\
&& + \frac{1}{2} l_1 (\bm{\nabla} \bm{Q} \vdots \bm{\nabla} \bm{Q} ) + \frac{1}{2} l_1 (\bm{\nabla} \cdot \bm{Q} \cdot \bm{\nabla} \cdot \bm{Q} ) + \frac{1}{2} l_3 \bm{Q}:\left( \nabla \bm{Q} : \nabla \bm{Q} \right) \nonumber\\ 
&&+ \frac{1}{2} b_1 \left|\bm{\nabla} \Psi\right|^2 + \frac{1}{4} b_2 \left|\nabla^2 \Psi\right|^2
\end{eqnarray}
\begin{eqnarray} \label{eq:free_energy_heterogenous_coeffs}
a & = & a_0 (T - T_{NI}) \nonumber \\
\alpha & = & \alpha_0 (T - T_{AI})\nonumber 
\end{eqnarray}
where $f$ is the free energy density, $f_0$ is the free energy density of the isotropic phase, terms 1-5 are the bulk contributions to the free energy, terms 6-7 are couplings of nematic and smectic order; both the bulk order and coupling of the nematic director and smectic density-wave vector, respectively.  Terms 8-10/11-12 are the nematic/smectic elastic contributions to the free energy.  $T$ is temperature, $T_{NI}$/$T_{AI}$ are the hypothetical second order transition temperatures for isotropic/nematic and isotropic/smectic-A mesophase transitions (refer to \cite{Coles1979a} for more detail), and the remaining constants are phenomenological parameters.

The Landau-Ginzburg time-dependent formulation \cite{Barbero2000} is used to capture the dynamics of the phase transition.  Due to the higher order derivative term in the free energy functional, a higher order functional derivative must be used.    Additionally, in order to utilise standard numerical solution techniques, the complex order parameter equation (\ref{eqsmec_order_param}) is separated into its real and imaginary contributions \cite{Ambrozic2004}:
\begin{equation} \label{eqcomplex_order_split}
\Psi = A+Bi
\end{equation}
The general form of the time-dependent formulation is as follows \cite{Barbero2000}:
\begin{eqnarray} \label{eq:landau_ginz}
\left(\begin{array}{c}
 \frac{\partial \bm{Q}}{\partial t}
\\ \frac{\partial A}{\partial t}
\\ \frac{\partial B}{\partial t} 
\end{array}\right)
&=& 
\left(\begin{array}{c c c} 
\frac{1}{\mu_n} & 0 & 0\\ 
0 & \frac{1}{\mu_S} & 0\\ 
0 & 0 & \frac{1}{\mu_S}\end{array} \right)
\left(\begin{array}{c} -\frac{\delta F}{\delta \bm{Q}}\\ 
-\frac{\delta F}{\delta A}\\ 
-\frac{\delta F}{\delta B} \end{array}\right)\\
F &=& \int_V f dV
\end{eqnarray}
where $\mu_r$/$\mu_s$ is the rotational/smectic viscosity, and $V$ the volume.  As previously mentioned, a higher order functional derivative must be used due to the second-derivative term in the free energy equation (\ref{eq:free_energy_heterogeneous}):
\begin{equation} \label{eqpdes}
\frac{\delta F}{\delta \theta} = \frac{\partial f}{\partial \theta} - \frac{\partial}{\partial x_i}\left(\frac{\partial f}{\partial \frac{\partial \theta}{\partial x_i}} \right) + \frac{\partial}{\partial x_i}\frac{\partial}{\partial x_j}\left(\frac{\partial f}{\partial \frac{\partial^2 \theta}{\partial x_i \partial x_j}} \right)
\end{equation}
where $\theta$ corresponds to the order parameter.

Substituting equation (\ref{eqcomplex_order_split}), the free energy (\ref{eq:free_energy_heterogeneous}), and high order functional derivative (\ref{eqpdes}) into the time-dependent formulation (\ref{eq:landau_ginz}) yields the closed set of model equations:
\begin{eqnarray} \label{eqthemodel}
\frac{\partial \bm{Q}}{\partial t} &=& -\left[ a^* \bm{Q} - b^* \left( \bm{Q} \cdot \bm{Q} \right)^{ST} + c^* \left( \bm{Q}:\bm{Q}\right)\bm{Q} - \delta^* \left( A^2 +B^2\right) \bm{Q} - \frac{1}{2} e^* \left( \bm{\nabla}A\bm{\nabla}A + \bm{\nabla}B\bm{\nabla}B \right)^{ST}\right] \nonumber\\
&&+ \bm{\nabla} \cdot \left( l_1^* \bm{\nabla Q} \right) \nonumber\\
\mu^* \frac{\partial A}{\partial t} &=& -\left[ \alpha^* A + \beta^* \left( A^2 + B^2 \right)A - \delta^* A \left(\bm{Q}:\bm{Q}\right)\right] +\bm{\nabla} \cdot \left[  b_1^* \bm{\nabla}A - e^* \bm{Q} \cdot \bm{\nabla} A - \frac{1}{2} b_2^* \bm{\nabla} \left(\nabla^2 A\right) \right]\nonumber\\ 
\mu^* \frac{\partial B}{\partial t} &=& -\left[ \alpha^* B + \beta^* \left( A^2  B^2 \right)B - \delta^* B \left(\bm{Q}:\bm{Q}\right)\right] + \bm{\nabla} \cdot \left[  b_1^* \bm{\nabla}B - e^* \bm{Q} \cdot \bm{\nabla} B - \frac{1}{2} b_2^* \bm{\nabla} \left(\nabla^2 B\right) \right]\nonumber\\ 
&&
\end{eqnarray}
where the asterisk denotes an nondimensionalized value, the superscript $ST$ denotes the symmetric/traceless portion of a tensor, and $\mu^*$ is the ratio of the smectic and rotational viscosities.  The nondimensionalized model parameters are as follows:
\begin{align}
a^* & = \frac{a_0 \overset{-}{T}}{\alpha_0}& b^* &= \frac{b}{\alpha_0 \Delta T}& c^* & =\frac{c}{\alpha_0 \Delta T} &\alpha^* & = \overset{-}{T}-1 & \beta^* &= \frac{\beta}{\alpha_0 \Delta T}& \delta^* & =\frac{\delta}{\alpha_0 \Delta T}& \nonumber\\
b_1^* & = \frac{b_1}{l^2 \alpha_0 \Delta T}& b_2^* & = \frac{b_2}{l^4 \alpha_0 \Delta T}& e^* & = \frac{e}{l^2 \alpha_0 \Delta T}& l_1^* & = \frac{l_1}{l^2 \alpha_0 \Delta T}& \nonumber \\
\mu^* &= \frac{\mu_s}{\mu_n}& \tau & = \frac{\mu_n}{\alpha_0 \Delta T}& \overset{-}{T}& = \frac{T - T_{ni}}{\Delta T}& \Delta T &= T_{AI}-T_{NI}&
\end{align}
where $l$ is the simulation-specific imposed length scale.

\subsection{Simulation method and conditions}

\begin{figure}
\begin{center}
\includegraphics[width=5cm]{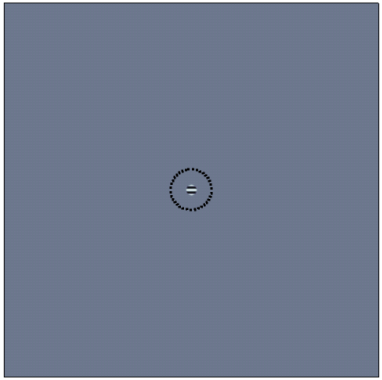}
\caption{Surface plot of $Re(\Psi)$ over the computational domain with the initial spherulite condition indicated by stippled circular region (assumed homogeneously oriented smectic-A nucleus). White/black corresponds to the maximum/minimum of $Re(\Psi)$ and the imposed length scale is $l=2.93\times 10^{-1} \mu m$.  The material parameters and phenomenological coefficients, based upon 12CB, are $T_{NI}=322.85K$, $T_{AI}=330.5K$, $a_0= 2\times10^5\frac{J}{m^3 K}$, $b=2.823\times10^7\frac{J}{m^3}$, $c=1.972\times10^7\frac{J}{m^3}$, $\alpha_0=1.903\times10^6\frac{J}{m^3 K}$, $\beta=3.956\times10^8\frac{J}{m^3}$, $\delta=9.792\times10^6\frac{J}{m^3}$, $e=1.938\times10^{-11}pN$, $l_1=1\times10^{-12}pN$, $b_1=1\times10^{-12}pN$, $b_2=3.334\times10^{-30}Jm$, $\mu_N = 8.4\times10^{-2}\frac{N \times s}{m^2}$, and the ratio of the rotational and diffusional viscosities used was $\frac{\mu_S}{\mu_N}=25$. \label{fig:initcond}}
\end{center}
\end{figure}

A square computational domain with an imposed length scale of $l=2.93\times 10^{-1} \mu m$ (approximately 75 smectic layers at $330 K$ \cite{Urban2005}, see figure \ref{fig:12CB}) was used in two separate simulations with isotropic nematic elasticity (no preferred interfacial anchoring $l_1 > 0$, $l_2,l_3 = 0$) and equal bend/splay nematic elasticity (preferred planar interfacial anchoring $l_1,l_2 > 0$, $l_3 = 0$).  Referring to figure \ref{fig:initcond}, Neumann boundary conditions were used to simulate bulk conditions.  The initial condition for both simulations was a smectic-A spherulite in an initially homogeneous layer configuration (see figure \ref{fig:initcond}).  The radius of the spherulite was initially set to $r_0=4.0nm$.  The initial value used for $S$, $\psi$, and the layer spacing correspond to the homogeneous values at $T=330K$, determined from the computed phase diagram (figure \ref{fig:12CB}).  The Heaviside step function was used to generate the initial spherulite.  The constraint that the spherulite does not impinge on the domain boundaries was verified post-simulation.

A commercial package, Comsol Multiphysics, was used to solve the time-dependent model (\ref{eqthemodel}).  Quadratic Lagrange basis functions were used for the Q-tensor variables and quartic Hermite basis functions used for the complex order parameter components.  Standard numerical techniques were utilised to ensure convergence and stability of the solution.  This platform does not support adaptive mesh refinement, thus a uniform mesh was used with a density of approximately $14.8$ nodes/$nm^2$.  Previous simulations using this model and numerical method have shown good agreement with both past experimental and theoretical findings \cite{Abukhdeir2008,Abukhdeir2008a}.  Additionally, exhaustive past work using this numerical method and the Landau-de Gennes model for the first-order isotropic/nematic phase transition \cite{Wincure2006,Wincure2007,Wincure2007a,Wincure2007b} has served to further validate this simulation approach.

\subsection{Multidimensional computation}

\begin{figure}
\begin{center}
\subfigure[]{}
\resizebox*{3cm}{!}{\includegraphics{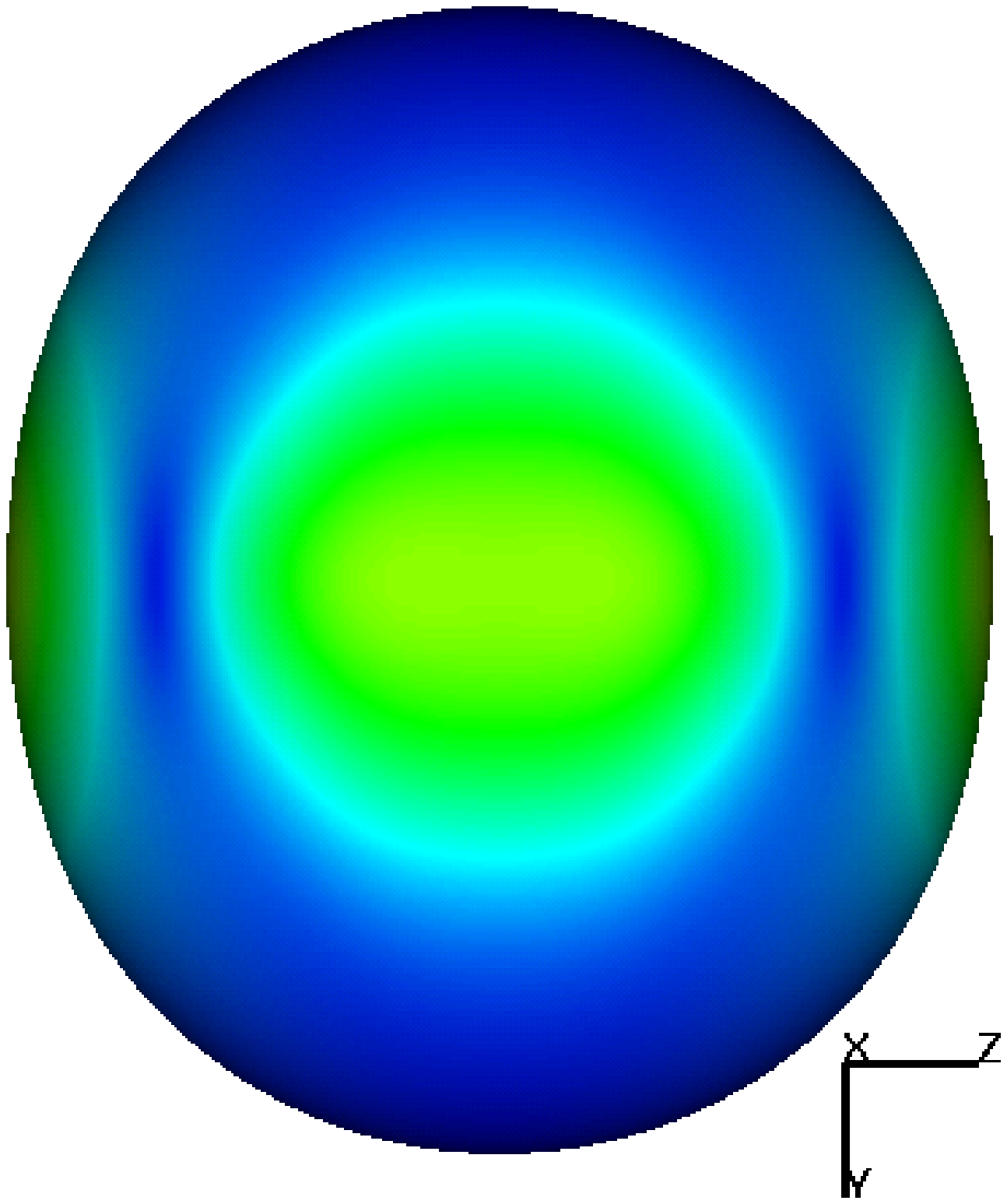}}
\subfigure[]{}
\resizebox*{3cm}{!}{\includegraphics{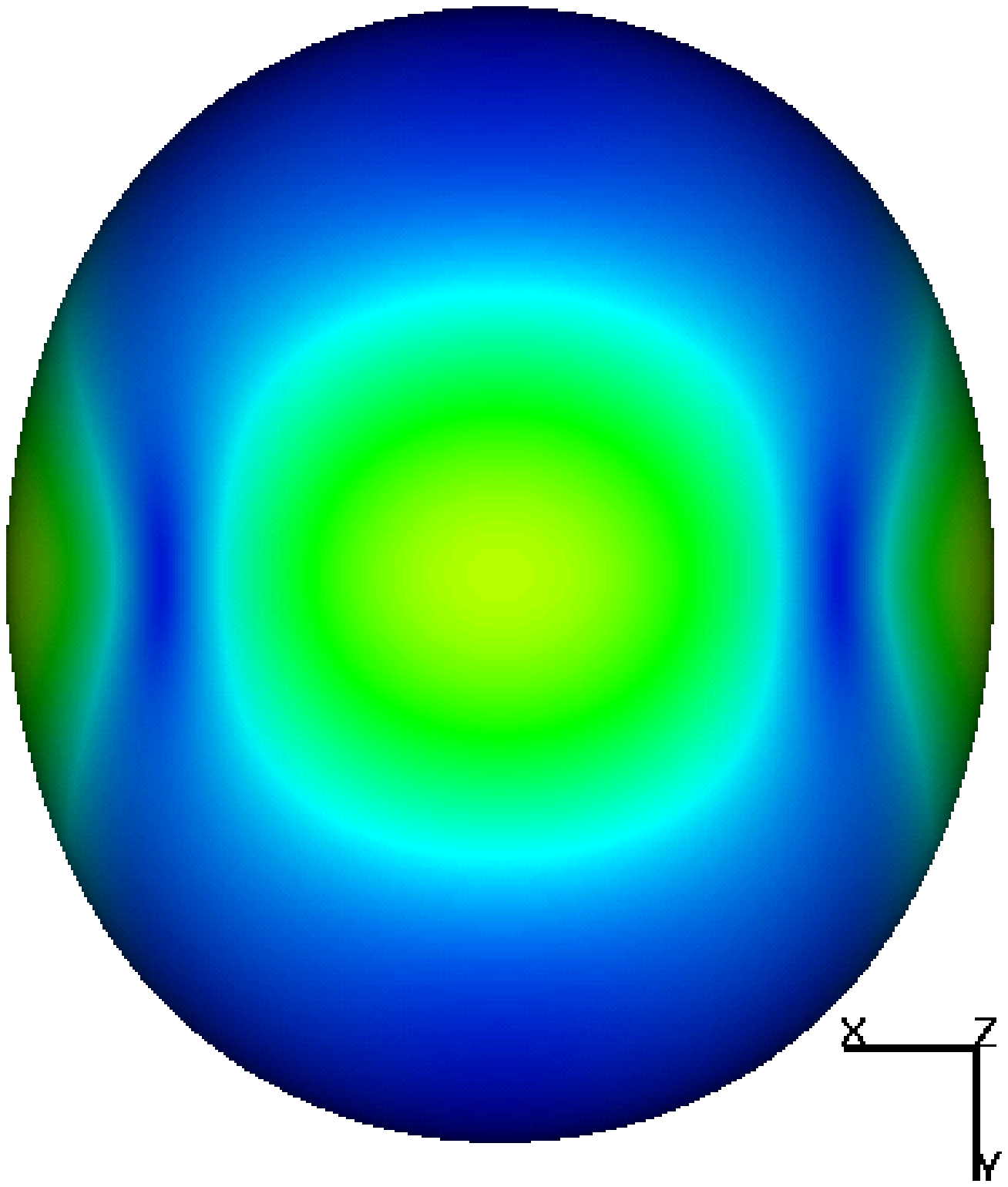}}
\subfigure[]{}
\resizebox*{3cm}{!}{\includegraphics{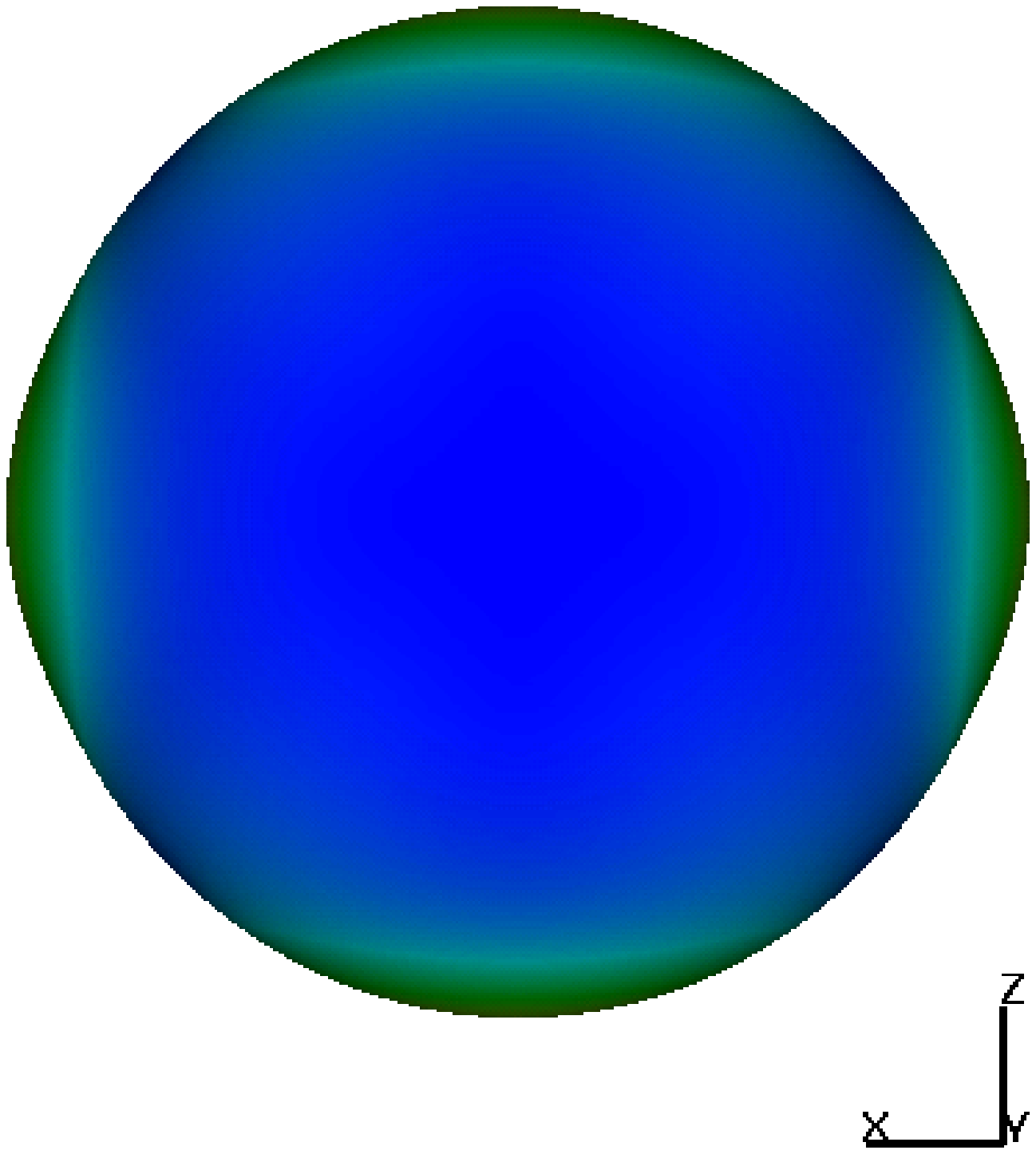}}

\subfigure[]{}
\resizebox*{3cm}{!}{\includegraphics{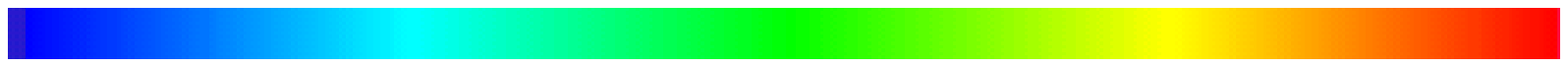}}

\caption{Results of growth of an initially homogeneous nematic spherulite in an isotropic matrix in three-dimensional simulation where the isosurface corresponds to $S=0.18$ and the surface shading corresponds to the biaxial order parameter $P$ (minimum/maximum $0/0.07$, see colour bar): (a) y/z-plane (b) x/y-plane (c) x/z-plane (d) colour bar for (a)-(c) ; phenomenological parameters were $\mu = 0.084 \frac{N \times s}{m^2}$, $T_{NI}=307.2 K$, $a_0=1.4\times 10^5 J/m^3\cdot K$, $b=1.8\times 10^7 J/m^3$, $c=3.6\times 10^6 J/m^3$, $l_1 =3.0\times 10^{-12} J/m$, $l_2=3.1\times 10^{-12} J/m$, $l_3 = 0.0\times 10^{-12} J/m$ \cite{Wincure2006}; horizontal length scale is $225 nm$, and a C++ finite element library, LibMesh 0.63 \cite{Kirk2006}, was used to develop the fully adaptive parallelised finite-element code. 
 \label{fig:2D3D}}
\end{center}
\end{figure}

Two fundamental challenges to research utilising numerical simulation are computational limitations and functionality of numerical routines.  As a result of the first challenge, the vast majority of simulation studies in the field of liquid crystals et al have been limited to one- and two-dimensions.  This is the case for the current study as well, but the symmetries and similarities between full three-dimensional simulation and two-dimensional simulation (of three dimensional phenomena) provide a strong motivation for and utility of obtaining these two-dimensional solutions.  A specific justification for this relationship can be found by comparing past simulation results of a growing nematic spherulite in two-dimensions \cite{Wincure2006,Wincure2007,Wincure2007a,Wincure2007b} with results of the same system (model, parameters, and initial conditions) in three-dimensions.  Figure \ref{fig:2D3D} shows results of three-dimensional simulation of the the growth of an initially homogeneous nematic spherulite based on two-dimensional studies by Wincure and Rey \cite{Wincure2006,Wincure2007,Wincure2007a,Wincure2007b}.  Figures \ref{fig:2D3D}a-c show three orthogonal views of an initially homogeneous nematic spherulite where the spherulite morphology in the y/z-plane (figure \ref{fig:2D3D}a) shows qualitative similarities to past two-dimensional results \cite{Wincure2006,Rey2007}.  Qualitatively, both the two-dimensional and three-dimensional initial conditions share symmetries in one orthogonal plane, thus the two-dimensional solution is qualitatively a subset of the three-dimensional result.  Based on this result, the two-dimensional smectic-A spherulite computations presented later are also expected to contain  signification features found in three-dimensional structures.

\section{Results and discussion}

Two different general types of post-nucleation growth are observed in growth processes of liquid crystals.  The shape-dynamic regime of growth from an initial nucleus, dominated by bulk elastic energy, involves shape and texturing dynamics as interfacial anchoring effects become important when the spherulite radius surpasses the characteristic length of liquid crystal ordering.  This shape-dynamics regime of growth transitions from the initial nucleus texture to a spherulite with shape and texture that minimises the overall free energy, allowing for constant growth to proceed.  Once this transition is complete, a self-similar growth regime is observed where spherulite texture and shape are independent of length scale and spherulite dimension scales with time $r \propto t^n$.  For the liquid crystal 5CB (pentyl-cyanobiphenyl), which exhibits an isotropic/nematic transition, the self-similar growth regimes for initially homogeneous spherulites were found to range from time-scales of approximately $80 \mu s$ and length scales of approximately $1.5 \mu m$ depending on temperature quench depth \cite{Wincure2007,Wincure2007a}.

\begin{figure}
\begin{center}
\subfigure[]{}
\resizebox*{3.3cm}{!}{\includegraphics{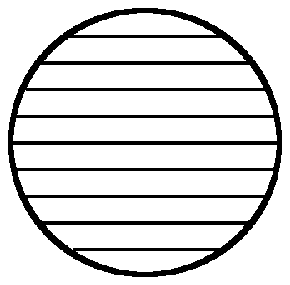}}
\subfigure[]{}
\resizebox*{3.3cm}{!}{\includegraphics{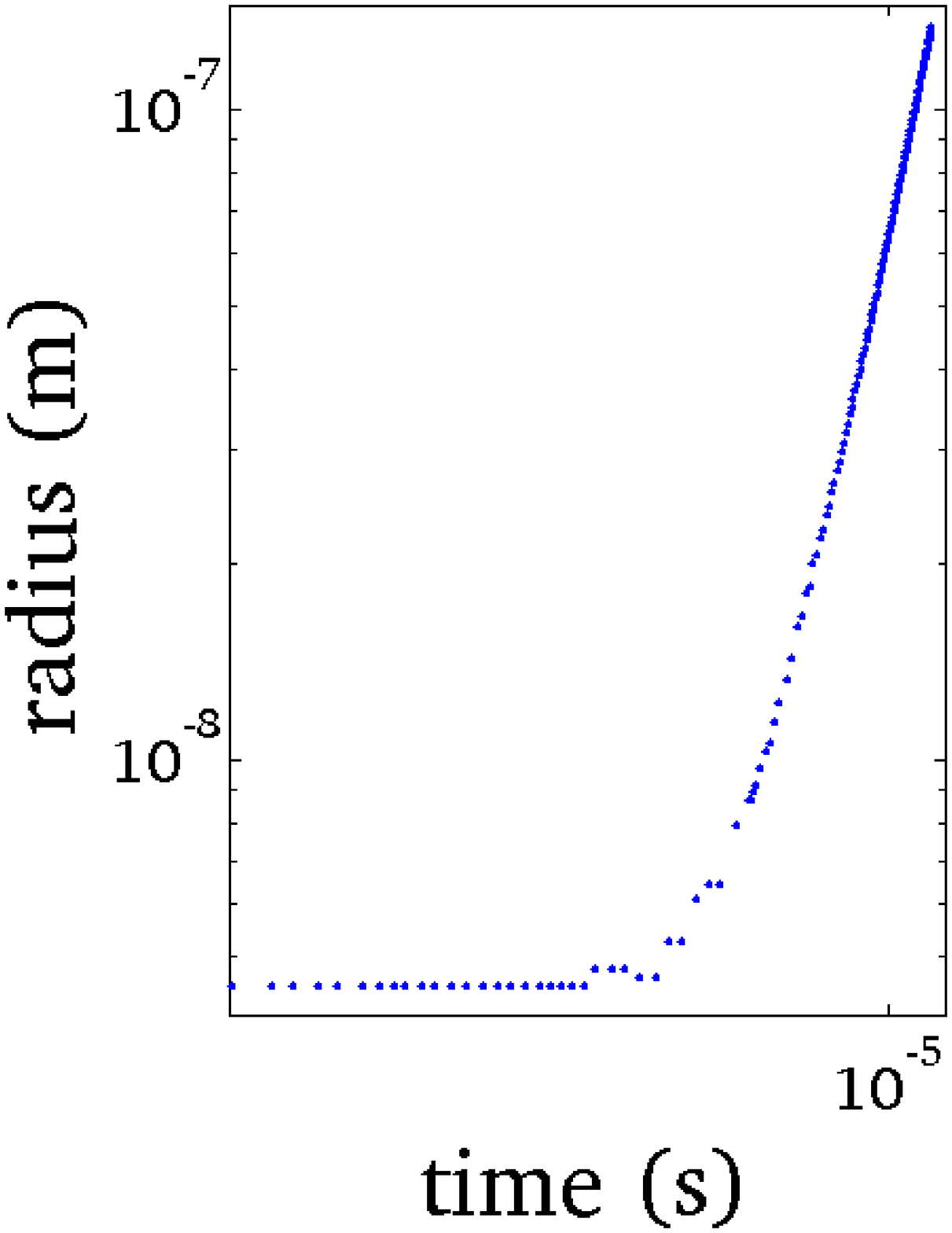}}
\subfigure[]{}
\resizebox*{3.3cm}{!}{\includegraphics{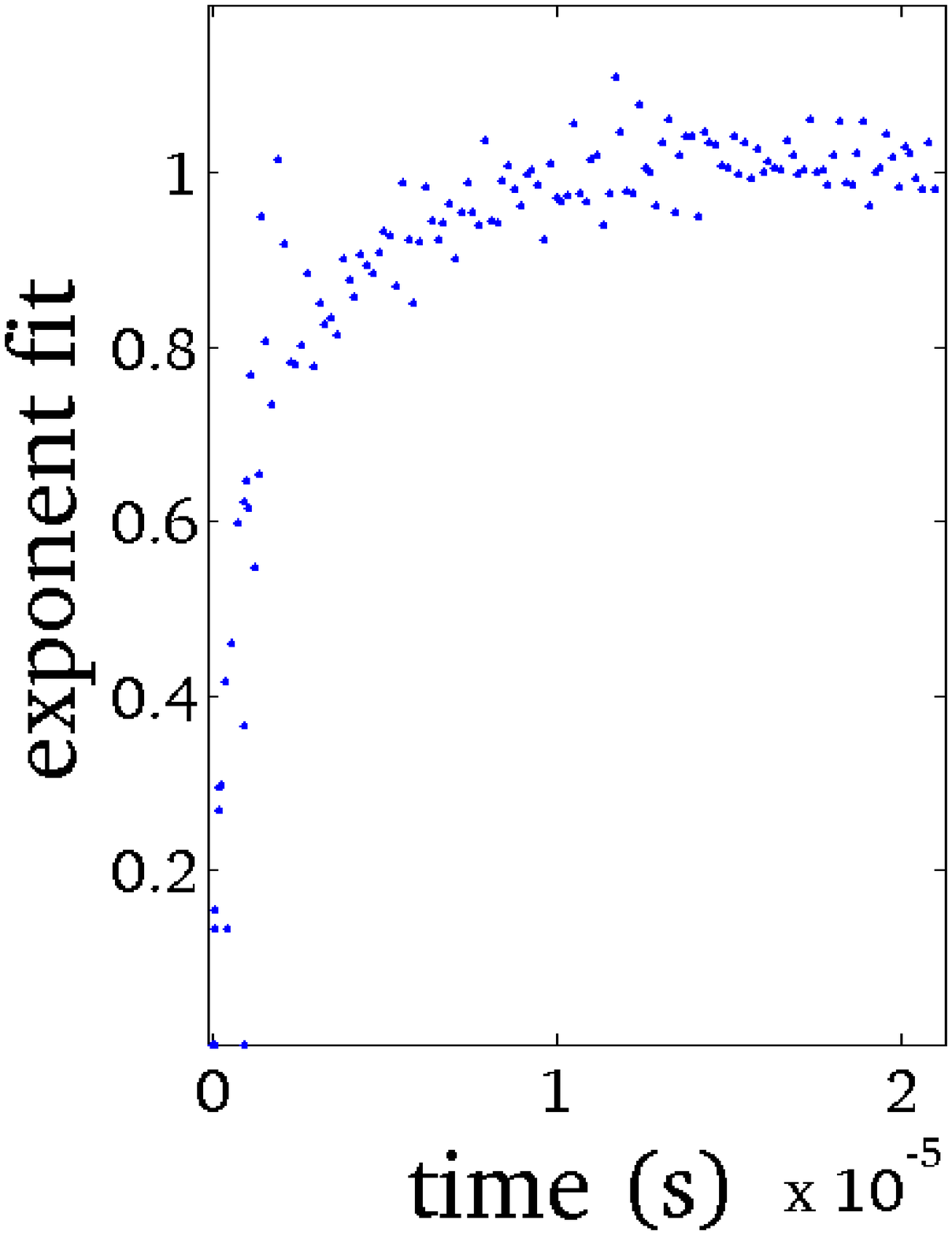}}

\caption{(a) Schematic of the Wulf construction result for the shape of a growing smectic-A spherulite with no preferred interfacial anchoring and an initially homogeneous bulk texture (b) log-log plot of the spherulite (major axis) radius versus time for the isotropic nematic elastic case (c) power-law fit of the spherulite (major axis) radius for the isotropic nematic elastic case. \label{fig:wulf}}
\end{center}
\end{figure}

Past work focused on an initially radially oriented nucleus \cite{Abukhdeir2008c}, where the initial nucleus shape has homogeneous interface conditions and the self-similar growth transition involves spherulite core dynamics only.  A homogeneously oriented nucleus has interfacial heterogeneities which result in bulk texture dynamics minimising total free energy, depending on the existence of preferred anchoring at the isotropic/smectic-A interface.  This results in a prolonged shape-dynamic growth regime which has been found to involve defect shedding in nematic cases \cite{Wincure2007a}.  The complex smectic ``batonnet'' structures composed of curvature defects first observed by Friedel and Grandjean \cite{Friedel1910,Dierking2003,Dierking2003a} have been attributed to growth from initially homogeneous smectic-A nuclei \cite{Fournier1991}.  Past approaches to studying these types of growth processes have involved highly simplified shape equations that take into account approximations of anisotropic interfacial anchoring energy and bulk energy contributions.  The simplest approach is that of the Wulf-construction \cite{Virga1994a} which determines surface shape by minimising the sum of total interfacial energy and an ideal undistorted bulk contribution:
\begin{equation} 
F = \int_A \gamma (\bm{r}) dA + \alpha V
\end{equation}
where $F$ is the total free energy of the spherulite, $\gamma$ is the interfacial tension (a function of position $\bm{r}$), and $\alpha$ the free energy density of the spherulite bulk.  While this approximation is suitable for crystal growth where a homogeneous bulk texture is a valid assumption, in the case of liquid crystals the occurrence of bulk elastic distortions and complexity of interfacial anchoring energies requires more rigorous approaches.  However, the Wulf approximation is a convenient starting point for analysis of the first simulation case with isotropic nematic elasticity.  In the absence of smectic-A ordering, the nematic contribution to the interfacial energy, neglecting curvature and biaxiality, is \cite{Yokoyama1997}:
\begin{equation} \label{eqn:int_tens}
\gamma=\frac{b^3\sqrt{3l_1+l_2/2+3l_2(\bm{n\cdot k})^2/2}}{486c^{5/3}}
\end{equation}
where $\bm{k}$ is the unit vector normal to the interface.  Equation \ref{eqn:int_tens} shows that in the case of nematic elastic isotropy ($l_2 = l_3 = 0$), the nematic contribution to the interfacial tension is approximately isotropic and there is no preferred interfacial anchoring of the nematic director.  Based upon this premise, the Wulf construction prediction for this situation should be valid in that interface anchoring conditions do not frustrate the bulk.  Figure \ref{fig:wulf}a shows the layer morphology of this scenario where a spherulite grows with a homogeneous texture and perfectly circular scale-independent shape.

Figures \ref{fig:wulf}b-c show simulation results of the long axis and its power law fit versus time for the isotropic nematic elastic case.  These results indicate that a transition from shape-dynamic to self-similar growth occurs on the order of $1 \mu s$ after nucleation.  These results are in agreement with past work on initially textured smectic-A spherulite growth \cite{Abukhdeir2008c} where in both cases the bulk elasticity is dominant over interfacial energy.  As previously mentioned, similar studies of nematic spherulite growth show that the transition from shape-dynamic to self-similar regimes occurs on a time scale on the order of $80 \mu s$ \cite{Abukhdeir2008b} due to the relatively lower magnitude of bulk elasticity of the nematic phase compared to interfacial anchoring energy.  An important conclusion from past work on nematic spherulite growth, that multiple shape-dynamic/self-similar regimes occur pre/post-shedding of defects \cite{Wincure2007}, which implies that similar phenomena could occur for smectic-A phase-ordering at length scales computationally unavailable.  In the smectic-A phase, disclination shedding events at the isotropic/smectic-A interface in conjunction with bulk texturing, could result in curvature defects such as focal conic domains observed in smectic-A batonnets \cite{Friedel1910}.

\begin{figure}
\begin{center}
\subfigure[]{}
\resizebox*{3.3cm}{!}{\includegraphics{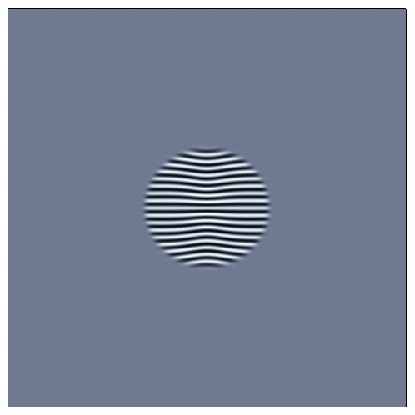}}
\subfigure[]{}
\resizebox*{3.3cm}{!}{\includegraphics{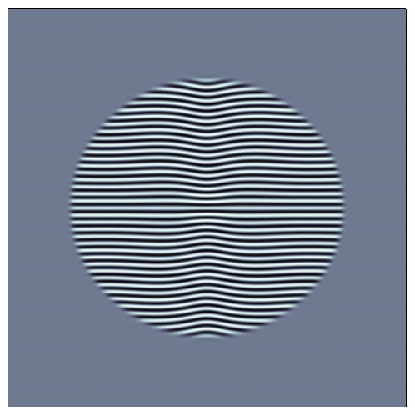}}
\subfigure[]{}
\resizebox*{3.3cm}{!}{\includegraphics{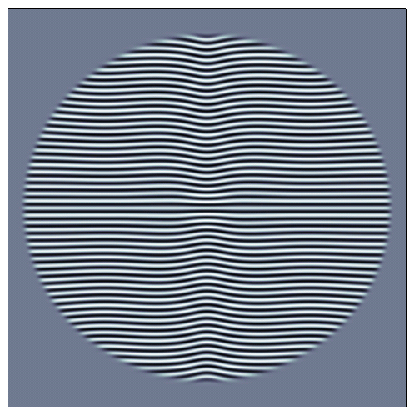}}

\subfigure[]{}
\resizebox*{3.3cm}{!}{\includegraphics{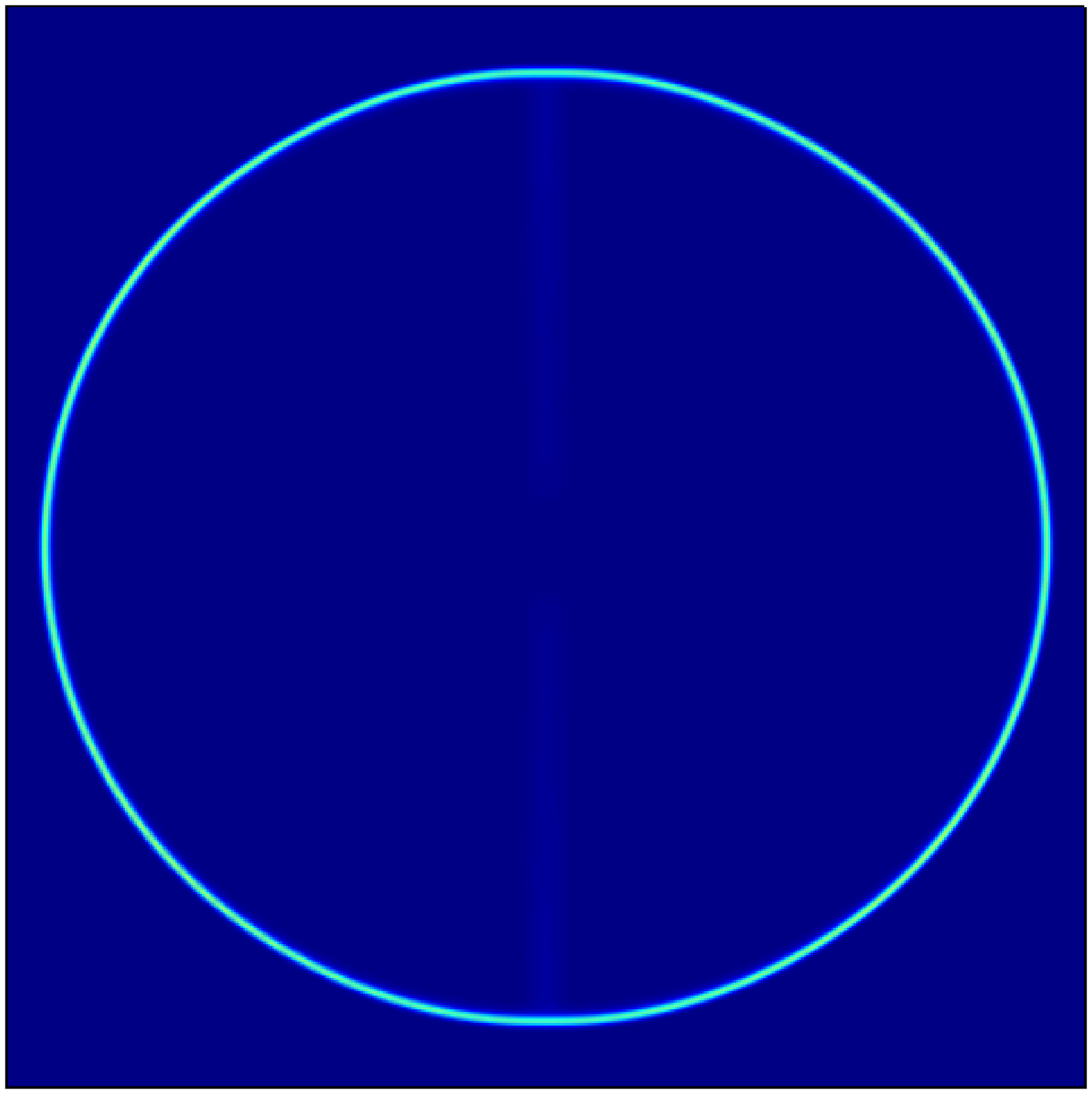}}
\subfigure[]{}
\resizebox*{3.3cm}{!}{\includegraphics{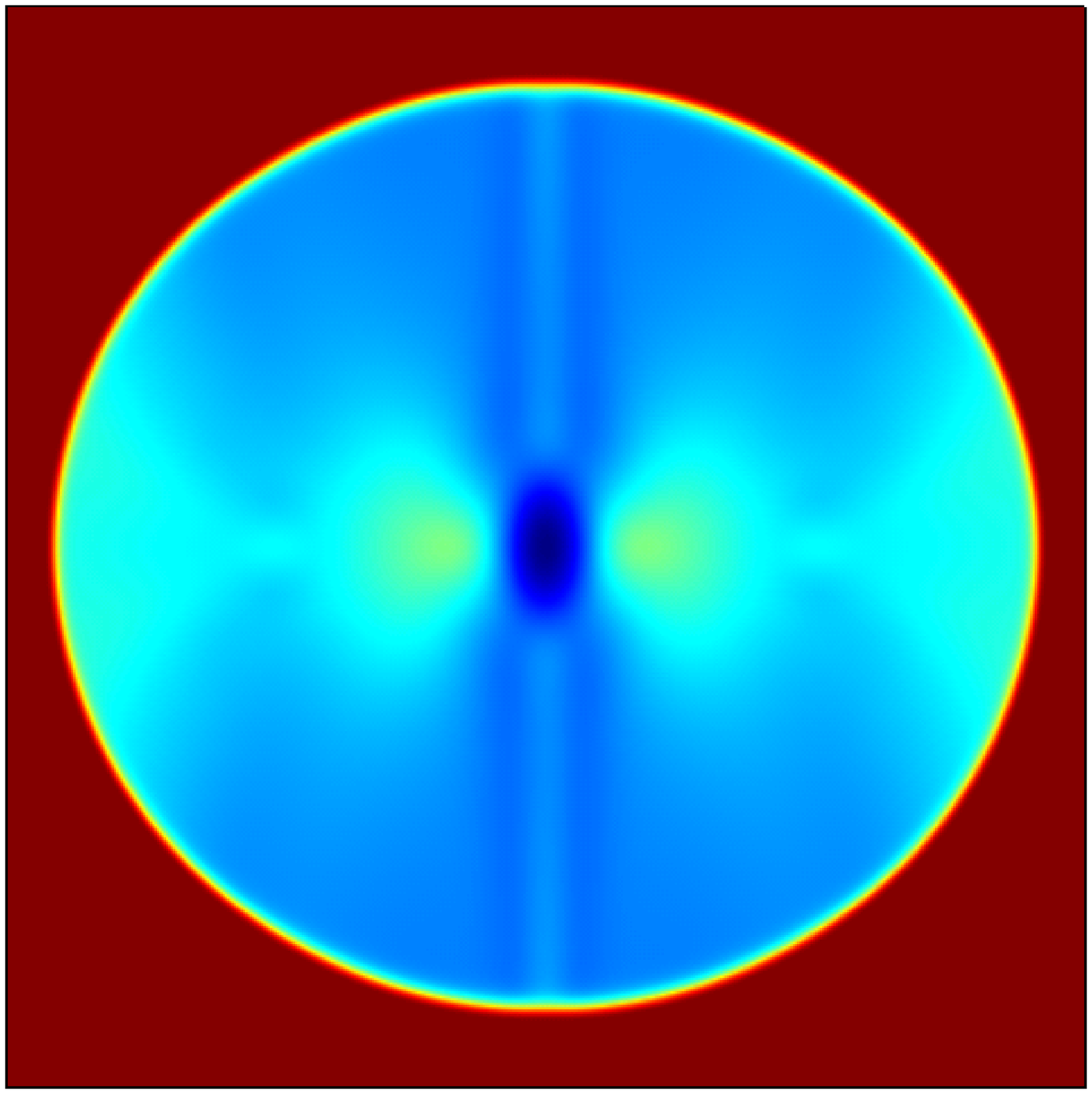}}
\subfigure[]{}
\resizebox*{3cm}{!}{\includegraphics{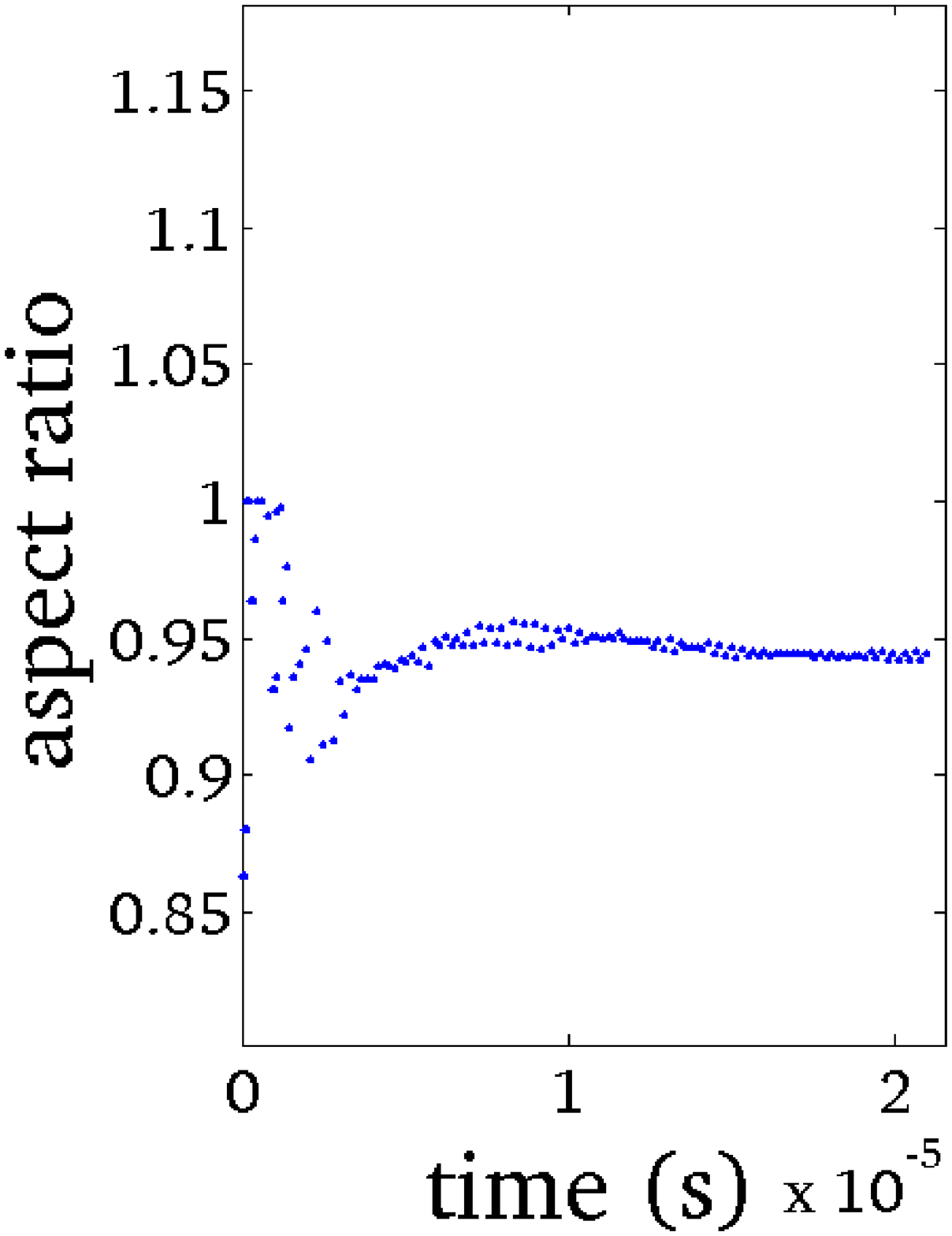}}

\subfigure[]{}
\resizebox*{3cm}{!}{\includegraphics{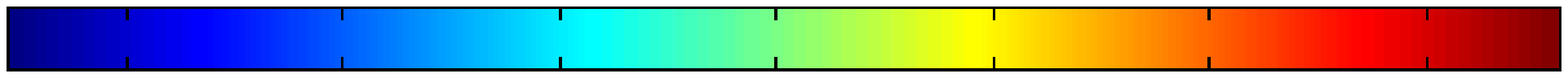}}

\caption{Simulation results for the isotropic nematic elastic case where the surface corresponds to Re($\Psi$) (minimum/maximum $-0.18/0.18$ corresponding to black/white) for simulations times (a) $6.7 \mu s$ (b) $15.0 \mu s$ (c) $21.0 \mu s$ (d) surface plot of the nematic elastic contribution (see equation \ref{eqn:nem_elast1}, minimum/maximum $0.0/2.151\times 10^4 J/m^3$, see colour bar) (e) surface plot of the smectic-A elastic contribution (see equation \ref{eqn:sma_elast1}, minimum/maximum $0.0/2.151\times 10^4 J/m^3$) (f) aspect ratio of y-/x- spherulite axes versus time for the isotropic nematic elastic case (g) colour bar for (d)-(e). \label{fig:morph1}}
\end{center}
\end{figure}

Figures \ref{fig:morph1}a-c shows the transient morphology of the simulation results for the isotropic nematic elastic case which does confirm that there is no substantial preferred interfacial anchoring from nematic elasticity (figure \ref{fig:morph1}d):
\begin{equation} \label{eqn:nem_elast1}
f_{Ne} = \frac{1}{2} l_1 (\bm{\nabla} \bm{Q})^2
\end{equation}
 However, the prediction of the Wulf construction (figure \ref{fig:wulf}a) is not found to be valid due to presence texturing in the bulk of the spherulite.  An undulation instability \cite{Chandrasekhar1992} is observed along the centreline of the spherulite parallel to the layer normal.  Figure \ref{fig:morph1}e shows the smectic-A elastic contributions:
\begin{equation} \label{eqn:sma_elast1}
f_{Ae} = \frac{1}{2} b_1 \left|\bm{\nabla} \Psi\right|^2 + \frac{1}{4} b_2 \left|\nabla^2 \Psi\right|^2 - \frac{1}{2} e \bm{Q}:\left(\bm{\nabla} \Psi\right)\left(\bm{\nabla} \Psi^*\right) 
\end{equation}
which indicates that this is a bulk phenomenon due to the gradient of the smectic order across the body of the spherulite.  As smectic order decreases approaching the interface with the isotropic phase, both the free energy penalty for layer dilation decreases (due to decreased smectic-A order) and the layer spacing increases (also due to decreased smectic-A order, see figure \ref{fig:12CB}).  The relationship was determined for a distortion-free smectic-A domain to be \cite{Abukhdeir2007}:
\begin{equation} \label{eqn:layerspace}
d_0 = 2 \pi \left(\frac{2 e S_A-3 b_1}{3 b_2} \right)^{-\frac{1}{2}} 
\end{equation}
where $d_0$ is the smectic-A layer spacing, $S_A$ is the nematic scalar order parameter in the smectic-A phase.  Due to symmetry, the core region of the spherulite has maximum smectic order and thus a minimum layer dilation.  This overall bulk layer dilation results in an undulation instability to dissipate higher energy layer expansion via low energy layer curvature \cite{Chandrasekhar1992}.

\begin{figure}
\begin{center}
\includegraphics[width=3cm]{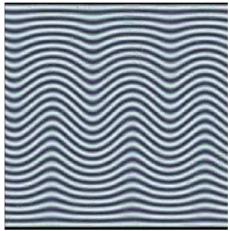}
\caption{Past simulation results, using the same material parameters in this work (see figure \ref{fig:initcond}), of undulation instabilities in a two-dimensional thin film geometry where the surface corresponds to Re($\Psi$).  The horizontal boundaries correspond to liquid crystal/solid interfaces and the vertical boundaries are periodic; the length scale of the domain is approximate $108 nm$ with an initially imposed undeformed texture of $25$ smectic-A layers ($4nm$ equilibrium layer spacing); adapted from ref. \cite{Abukhdeir2008}  \label{fig:undulation}}
\end{center}
\end{figure}

This layer dilation undulation instability, referred to as Helfrich-Huralt undulations \cite{Chandrasekhar1992,Ishikawa2001}, has been observed in films of lamellar systems including smectic-A and cholesteric liquid crystals under an external layer dilation force \cite{Chandrasekhar1992}.  Figure \ref{fig:undulation} shows a representative example of past results for the currently simulated model system in a two-dimensional thin film geometry with layer dilation imposed by perturbing the plate separation distance.  As derived using linearised lamellar elasticity \cite{deGennes1995}, the optimal wavelength for the undulation distortion in the thin film geometry (figure \ref{fig:undulation}) is approximately \cite{Chandrasekhar1992}:
\begin{equation}
\lambda_U = 2 \pi \left( \frac{\lambda L}{\pi} \right)^{0.5}
\end{equation}
where $\lambda$ is the characteristic length of the lamellar ordering (on the order of the layer spacing), and $\lambda_U$ is the characteristic length of the undulation instability, and $L$ is the externally imposed length scale.  Numerical simulation results shown in figure \ref{fig:morph1}e and the ideal layer spacing dependence equation \ref{eqn:layerspace} shows the source of the layer dilation driving the instability.  Past work studying a quasi-lamellar system, the cholesteric mesophase, has shown that cholesteric pitch gradients (similar to the smectic-A layer spacing) are imposed by the presence of the interface with the isotropic phase for this system as well \cite{Rey2000}.  These past results along with the current observations imply that a similar undulation instability should be seen in growing cholesteric spherulites, although experimental evidence of this phenomena has not been observed.  This can be explained based upon comparing magnitudes of $\lambda_U$ and spherulite radii for both cases, using estimates of their characteristic lengths.  For typical smectic-A liquid crystals, $\lambda$ is on the order of $nm$ which results in $\lambda_U < R$ in the self-similar regime.  Thus an undulation instability is predicted to be observed in this growth scenario as is confirmed with the present simulations.  For cholesteric liquid crystals, the characteristic length is on the order of $\mu m$ which is also on the order of the maximum spherulite size that has been observed experimentally.  Thus $\lambda_U \ge L$ which explains why the growth-induced undulation instability observed in the smectic-A system is not observed in the quasi-lamellar cholesteric liquid crystal.

An additional effect of this texture induced by the introduction of smectic-A order is that the spherulite shape deviates from the Wulf construction prediction of perfectly spherical growth.  Instead, the spherulite has a long axis parallel to the smectic layers in order to minimise the over spherulite area with high layer dilation.  The time evolution of the spherulite aspect ratio, shown in figure \ref{fig:morph1}f shows a convergence to a value slight below $1$ resulting from an equilibrium bulk texture without the presence of frustration from interfacial anchoring, but instead from the initial homogeneous texture and order parameter gradients induced by the presence of the interface.

\begin{figure}
\begin{center}
\subfigure[]{}
\resizebox*{3.3cm}{!}{\includegraphics{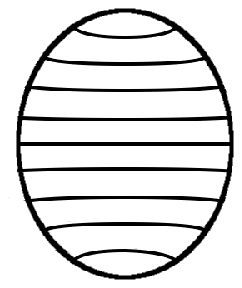}}
\subfigure[]{}
\resizebox*{3.3cm}{!}{\includegraphics{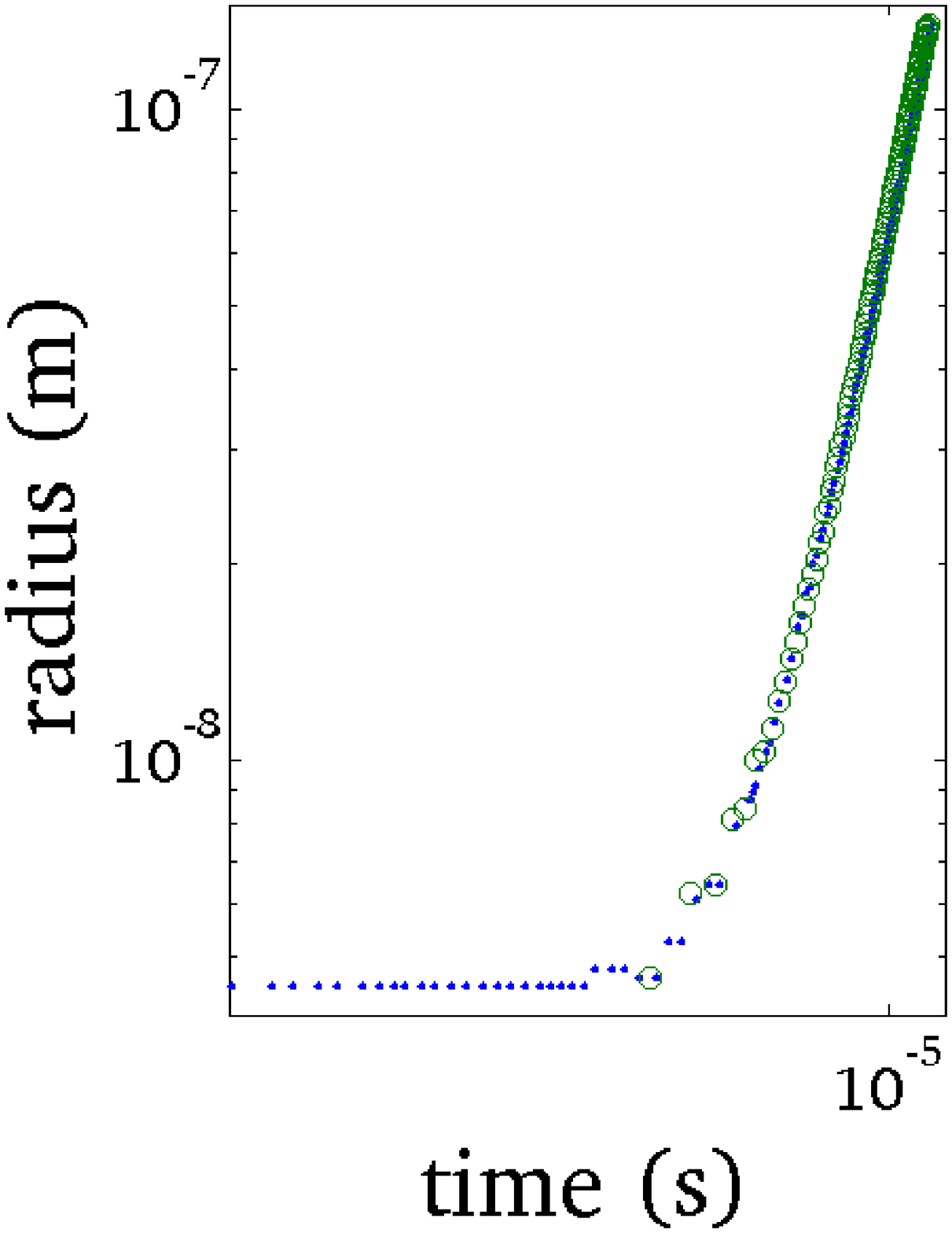}}
\subfigure[]{}
\resizebox*{3.3cm}{!}{\includegraphics{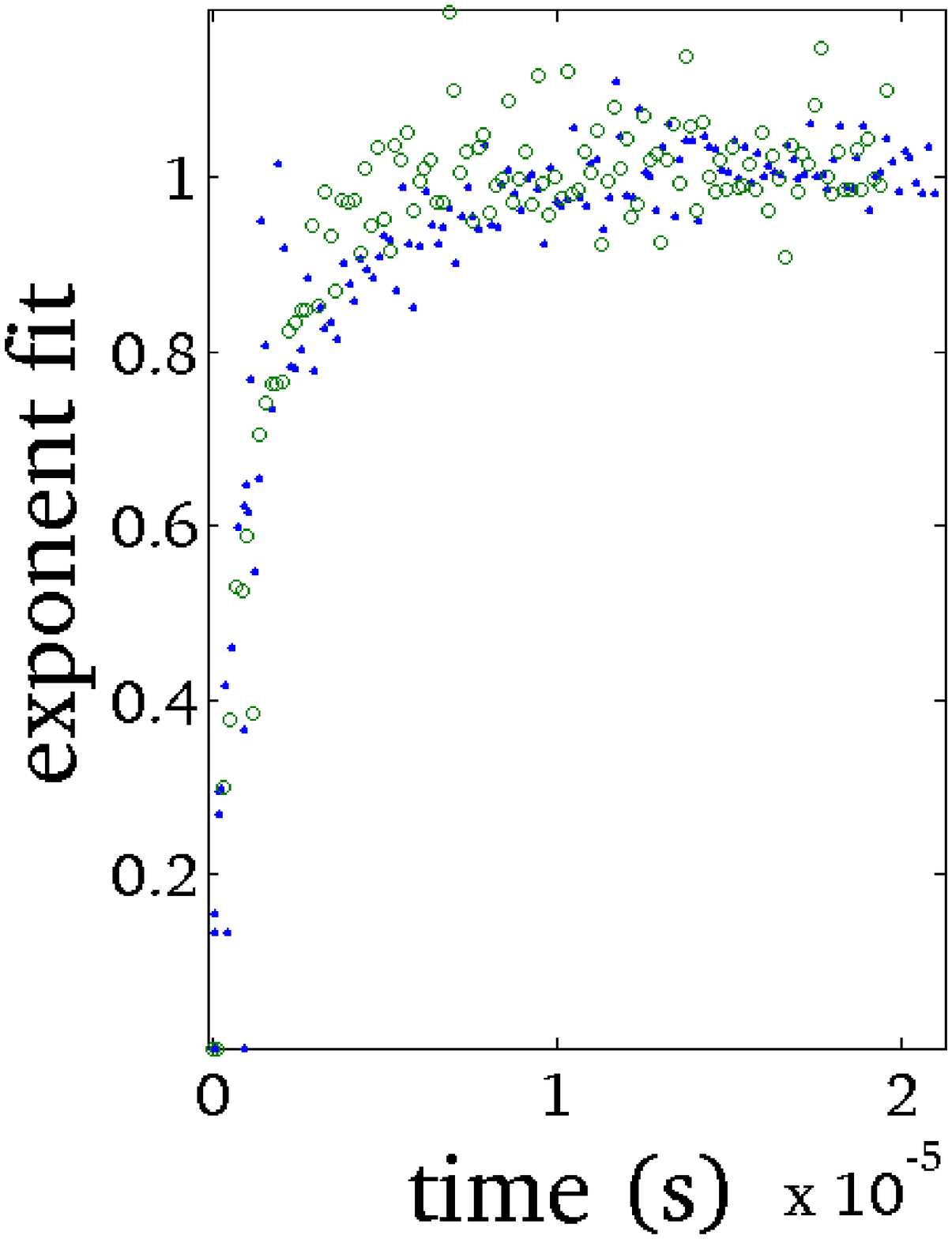}}

\caption{(a) spherulite shape/morphology predicted by Fournier and Durand's extension to the Wulf construction (based on figure 6b from ref. \cite{Fournier1991})(b) log-log plot of spherulite (major axis) radius versus time for the isotropic nematic elasticity case (c) power-law fit of spherulite (major axis) radius versus for the isotropic nematic elasticity case \label{fig:fournier}}
\end{center}
\end{figure}

For the equal splay-bend nematic case, equation \ref{eqn:int_tens} shows that the nematic contribution to the interfacial anchoring prefers a planar orientation to the isotropic interface (average orientational axis parallel to the interface).  This scenario is in agreement with experimental observations of preferred planar anchoring for the isotropic/smectic-A interface \cite{Fournier1991}.  In this case the Wulf-construction is not adequate in that frustration of the interfacial anchoring with the bulk texture will result in deviation from homogeneous bulk ordering.  A simple extension of Wulf's approach was made by Fournier and Durand which takes into account finite smectic-A elasticity \cite{Fournier1991}:
\begin{equation} \label{eqn:fourn1}
F = \int_A \gamma dA + \int_V f_d dV + \alpha V
\end{equation}
As with the Wulf-construction, minimisation of the total free energy of the spherulite $F$ predicts its shape, but an additional term is used to describe all possible discrepancies with respect to the ideal homogeneous bulk texture:
\begin{equation} \label{eqn:fourn2}
f_d(\mathbf{r}) = f_{el}(\mathbf{r}) + \delta f\{\psi(\mathbf{r} - \psi_{eq})\}
\end{equation}
where $f_{el}$ is the elastic free energy density (curvature/dilation) and the second term describe melting of smectic-A order due to defects.  Fournier and Durand determine a semi-quantitative approximate solution to equations \ref{eqn:fourn1}-\ref{eqn:fourn2} which predicts a relaxed configuration as shown in figure \ref{fig:fournier}a.  This solution determines a spherulite configuration where, to a first approximation, dilation, curvature, and interfacial anchoring energies are minimised.  Non-dilative configurations exist which involve complex curvature defects, or focal conics, in order to minimise total energy of the growing spherulite \cite{Fournier1991}.  Fournier and Durand determine a specific spherulite focal conic texture which results in a total free energy substantially lower than that predicted by the extended Wulf's method \cite{Fournier1991}.  This type of approach is limited in that predictions can be made, but transition mechanisms and alternate dilation-minimum modes are not energetically identified from a growth process.  Thus pure geometrical approaches, or those extended to account for dilation/curvature elasticity are not able to take into account an adequate subset of the physics involved in these growth processes.

\begin{figure}
\begin{center}
\subfigure[]{}
\resizebox*{3.3cm}{!}{\includegraphics{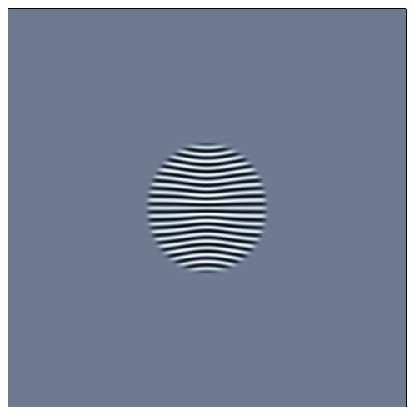}}
\subfigure[]{}
\resizebox*{3.3cm}{!}{\includegraphics{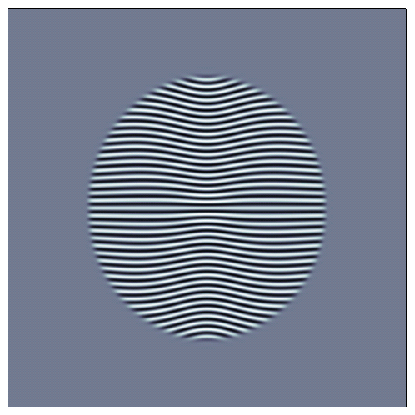}}
\subfigure[]{}
\resizebox*{3.3cm}{!}{\includegraphics{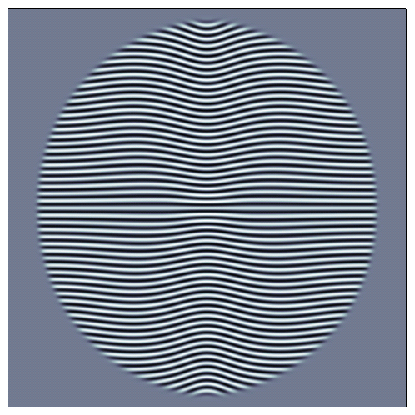}}

\subfigure[]{}
\resizebox*{3.3cm}{!}{\includegraphics{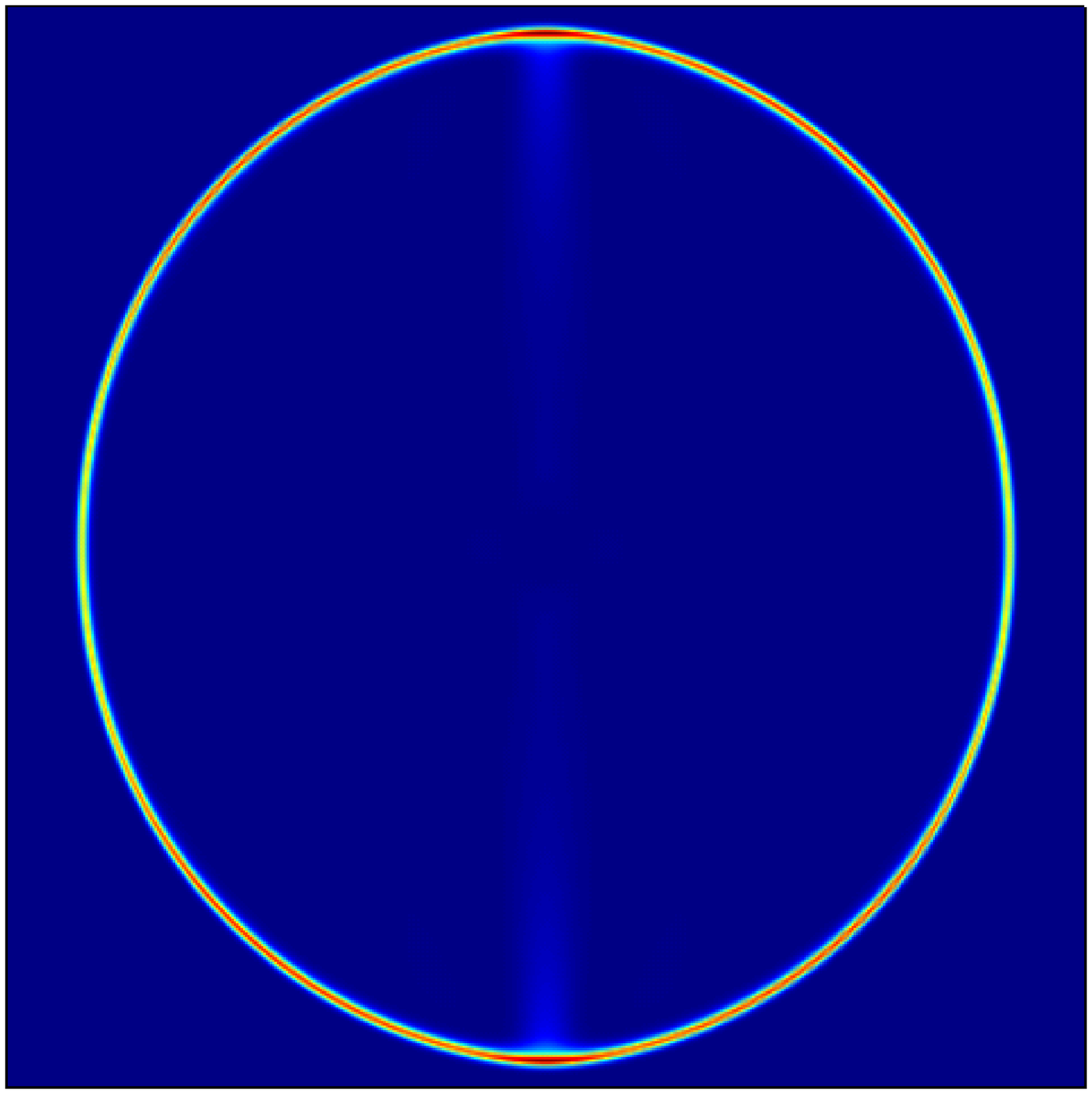}}
\subfigure[]{}
\resizebox*{3.3cm}{!}{\includegraphics{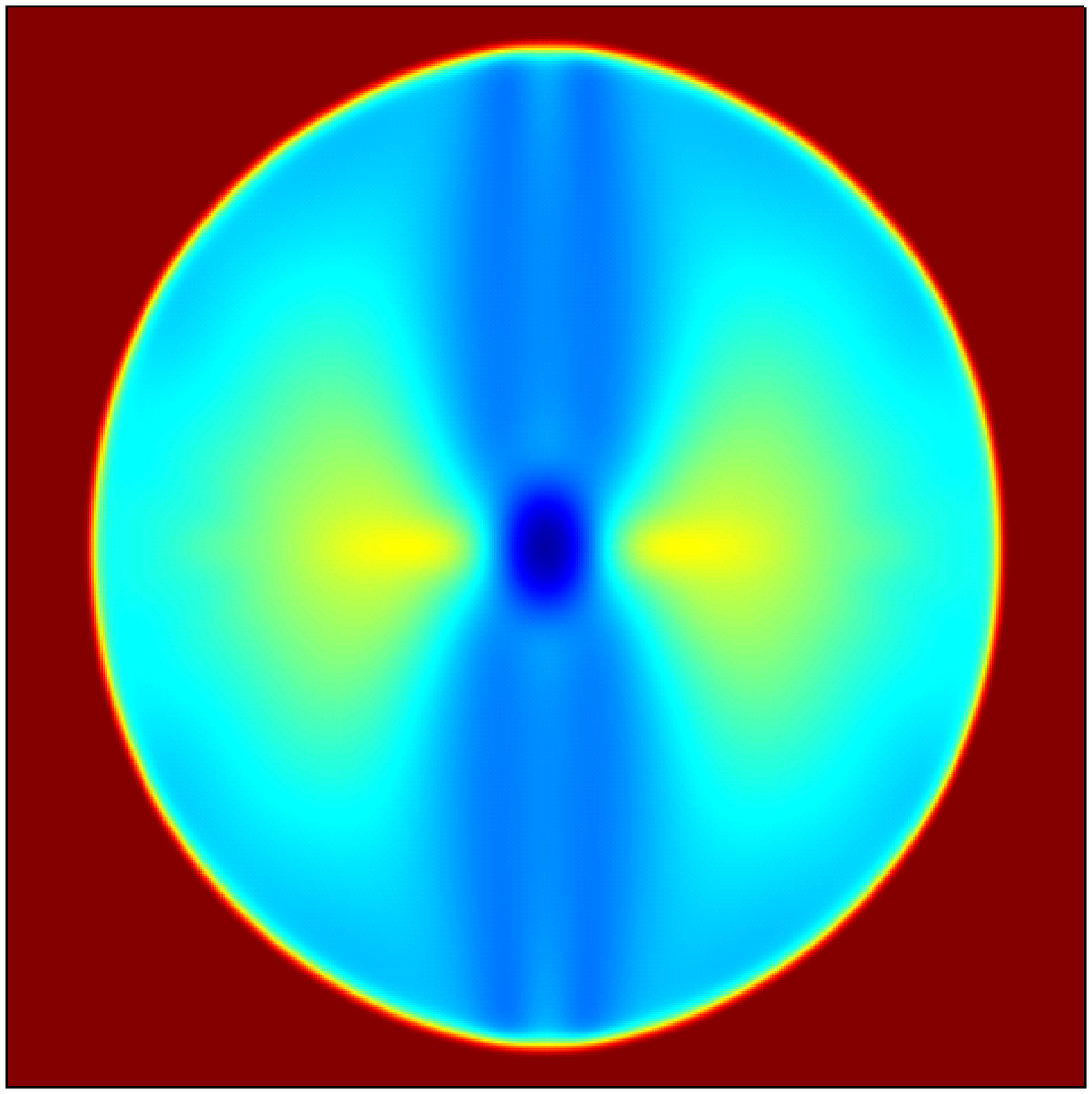}}
\subfigure[]{}
\resizebox*{3cm}{!}{\includegraphics{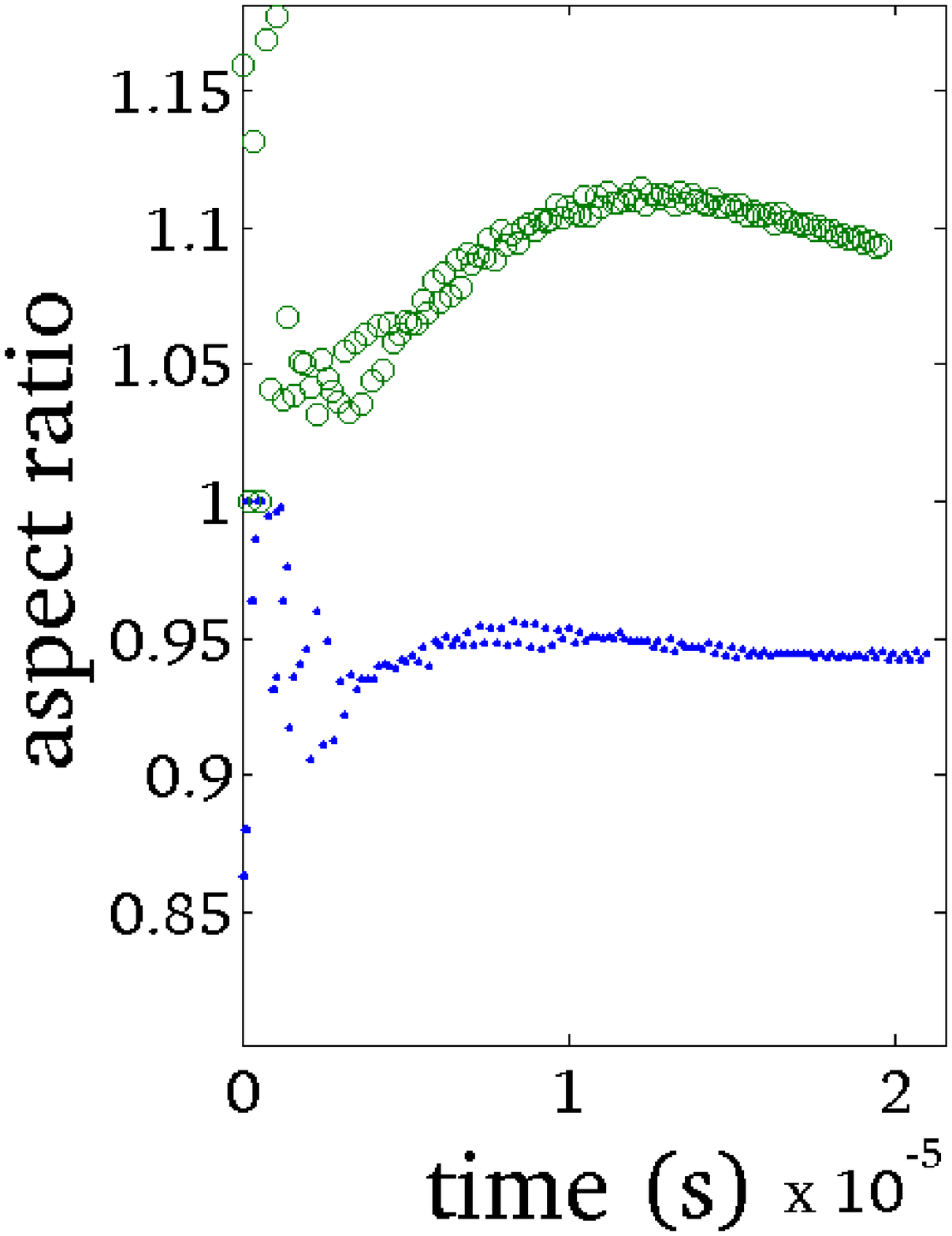}}

\subfigure[]{}
\resizebox*{3cm}{!}{\includegraphics{images/jet_colorbar}}

\caption{Simulation results for the equal nematic splay-bend case where the surface corresponds to Re($\Psi$) (minimum/maximum $-0.18/0.18$ corresponding to black/white) for simulations times (a) $6.4 \mu s$ (b) $13.7 \mu s$ (c) $19.6 \mu s$. (d) surface plot of the nematic elastic contribution (see equation (\ref{eqn:nem_elast2}), minimum/maximum $0.0/2.151\times 10^4 J/m^3$, see colour bar) (e) surface plot of the smectic-A elastic contribution (see equation (\ref{eqn:sma_elast1}), minimum/maximum $0.0/2.151\times 10^4 J/m^3$) (f) aspect ratio of y-/x spherulite axes versus time for both simulation cases (g) colour bar for (d)-(e). \label{fig:morph2}}
\end{center}
\end{figure}

Figures \ref{fig:fournier}b-c show simulation results of the long axis and power law fit versus time for the nematic elastic isotropy case superimposed on those results from the nematic elastic isotropy case.  The introduction of anisotropic interfacial tension due to the equal bend-splay elasticity (figure \ref{fig:morph1}d):
\begin{equation} \label{eqn:nem_elast2}
f_{Ne} = \frac{1}{2} l_1 (\bm{\nabla} \bm{Q} )^2 + \frac{1}{2} l_2 (\bm{\nabla} \cdot \bm{Q} )^2
\end{equation}
 results in little change in the growth kinetics compared to the previous case.  Figures \ref{fig:morph2}a-c show the dynamic morphology of the spherulite which is in good agreement with the shape prediction of Fournier and Durand (figure \ref{fig:fournier}a).  Again, the undulation instability is present in addition to interfacial heterogeneity from the preferred planar anchoring.  The additional smectic layer curvature resulting from the transition from homeotropic to planar anchoring (molecular axis perpendicular to the interface) at the spherulite poles also promotes the undulation instability resulting in an increased amplitude compared a spherulite of equal vertical radius in the previous nematic elastic case.

Figure \ref{fig:morph2}f shows that the evolution of the aspect ratio differs compared to the previous nematic elastic case, where now the aspect ratio decays as the spherulite grows from a maximum value following the shape-dynamic growth regime.  While the spherulite shape morphology and preferred anchoring of the isotropic/smectic-A interface are in agreement with past theoretical and experimental observations, respectively, the decay of the aspect ratio is not in-line with experimental observation of high-aspect ratio batonnets.  This decay while in the self-similar growth regime can be explained based upon a scaling theory derived from the studying shape and director-field transformation of nematic tactoids \cite{Prinsen2003,Prinsen2004}.  This approach uses a simplified nematic spherulite free energy which takes into account both interfacial energy coupled with bulk elastic anchoring as a function of the nematic director:
\begin{equation} \label{eqn:bipol1}
F = \tau \int_A \left(1 + \omega (\bm{q} \cdot \bm{n})^2 \right) dA + \frac{1}{2} K \int_V \left( (\bm{\nabla} \cdot \bm{n})^2 + \left[ \bm{n} \times (\bm{\nabla} \times \bm{n})\right]^2\right) dV
\end{equation}
where $\tau$ is interfacial tension, $\omega$ is a dimensionless anchoring strength, $\bm{q}$ is the surface normal (a function of position $\bm{r}$), $\bm{n}$ is the nematic director, and $K$ is the elastic constant (using an equal splay-bend assumption).  Assuming a fixed bipolar nematic texture (see figure \ref{fig:fournier}a where contour lines/smectic layers indicate the orientation of the vector perpendicular to the nematic director $\bm{n}$) and a fixed spherulite volume $V$, a scaling estimate for the aspect ratio of the spherulite obeys the following relationship to minimise total free energy:
\begin{equation} \label{eqn:bipol2}
\frac{R}{r} \approx K^{3/5}\tau^{-3/5}V^{-1/5}
\end{equation}
where $R$ is the major axis and $r$ is the minor axis.  Equation (\ref{eqn:bipol2}) predicts that the spherulite aspect ratio decreases with volume, as is found for the equal splay-bend case in figure \ref{fig:morph2}f.  This implies that as the spherulite radius grows the aspect ratio will converge to $1$, assuming no additional shape-dynamic events, which is not in agreement with observations of complex batonnet structures in isotropic/smectic-A transitions.  Thus it is expected that if computational resources allowed access to larger spherulite length scales, another shape-dynamics regime would be observed similar to defect shedding events seen during the growth of nematic spherulites \cite{Wincure2007,Wincure2007a}.

\section{Conclusions}

The study of the two-dimensional growth of an initially homogeneous smectic-A spherulite in an isotropic matrix was performed via modeling and simulation.  A mesoscale Landau-de Gennes type model was used that takes into account the coupling between orientational (nematic) and one-dimensional translational (smectic-A) order where layer spacing and the coupling of the nematic director and wave-vector are energetically imposed.  The effect of two different nematic elasticity conditions corresponding to isotropic and preferred planar interfacial anchoring were performed and compared to past experimental and theoretical studies:
\begin{itemize}
\item A growth-induced dilative layer undulation instability was observed unique to initially homogeneous smectic-A spherulite growth.  This instability was shown to be independent of interfacial anchoring and a result of gradients in the bulk order.
\item The use of nematic equal bend-splay elastic conditions was shown to be required to adequately model the preferred planar interfacial anchoring experimentally observed for the isotropic/smectic-A interface.
\item The use of nematic equal bend-splay elastic conditions was shown to result in a spherulite shape in agreement with past theoretical predictions for both smectic-A spherulites and similarly textured nematic spherulites.
\item An aspect ratio decay of the spherulite was observed in the self-similar growth regime, which implies that a second shape-dynamics regime follows at greater spherulite radius in order to agree with experimental observations of smectic-A batonnets with high aspect ratio.  A defect shedding shape-dynamics process is proposed, similar to that seen in growing nematic spherulites \cite{Wincure2007,Wincure2007a}, to transition the bipolar oriented smectic-A spherulite to one in the diverse set of batonnet structures composed of focal conic curvature defects.
\end{itemize}

This and past simulation work \cite{Abukhdeir2007,Abukhdeir2008,Abukhdeir2008a,Abukhdeir2008c} using the high-order Landau-de Gennes type phenomenological model of Mukherjee, Pleiner, and Brand \cite{deGennes1995,Mukherjee2001} has shown great promise for studying both the isotropic/smectic-A transition and the overall smectic-A mesophase.  The key aspects of this model, in addition to its phenomenological nature, are that energetic couplings are employed between orientational/translational (nematic/smectic-A) order including: bulk nematic and smectic-A order ($S/\psi$), average molecular axis and smectic layer normal ($\bm{n}/\bm{a}$), and bulk nematic/smectic-A order and layer spacing ($S/\psi/|\bm{a}|$).  These couplings and the use of full tensorial and complex order parameters capture a sub-set of the fundamental physics involved in the isotropic/smectic-A transition unavailable using more simplified approaches.

The main limitation, as with most modeling approaches, lies in the limits imposed computationally.  Cutting-edge scientific computing approaches such as fully adaptive parallelised finite element libraries (for example, LibMesh \cite{Kirk2006} and PETSc-FEM \cite{Storti2008}) maximize the impact of currently available computational resources.  Through the use of these existing numerical approaches, development of parallelised post-processing code, and the presently used high-order model, smectic-A phenomena at macroscopic lengths scales could be accessed to study diverse multi-scale growth phenomena such as the fascinating batonnet structures of Friedel and Grandjean \cite{Friedel1910,Dierking2003,Dierking2003a}.

\section*{Acknowledgements}

This work was supported by a grant from the Natural Science and Engineering Research Council of Canada.

\bibliographystyle{tLCT}
\bibliography{/home/nasser/nfs/references/references}

\begin{thebibliography}{43}
\providecommand{\natexlab}[1]{#1}

\bibitem[1]{Lockwood2008}
Lockwood, N.; Gupta, J.; Abbott, N. Self-assembly of amphiphiles, polymers and
  proteins at interfaces between thermotropic liquid crystals and aqueous
  phases.  {\em Surface Science Reports}  {\bf 2008}, {\em 63}, 255?293.

\bibitem[2]{Fisch2004}
Fisch, M.R. Liquid Crystals, Laptops and Life. In: {\em World Scientific Series
  in Contemporary Chemical Physics};  Vol. 23,   : , 2004.

\bibitem[3]{Rizvi2003}
Rizvi, T.Z. Liquid crystalline biopolymers: A new arena for liquid crystal
  Research.  {\em Journal of Molecular Liquids}  {\bf 2003}, {\em 106}, 43--53.

\bibitem[4]{Aldoroty1987}
Aldoroty, R.; Garty, N.; April, E. Donnan potentials from striated muscle
  liquid crystals. Lattice spacing dependence.  {\em Biophysical Journal}  {\bf
  1987}, {\em 51} (3), 371--381.

\bibitem[5]{Nakata2007}
Nakata, M.; Zanchetta, G.; Chapman, B.D.; Jones, C.D.; Cross, J.O.; Pindak, R.;
  et~al. End-to-End Stacking and Liquid Crystal Condensation of 6 to 20 Base
  Pair DNA Duplexes.  {\em Science}  {\bf 2007}, {\em 318} (5854), 1276--1279.

\bibitem[6]{Rey2007}
Rey, A.D. Capillary models for liquid crystal fibers, membranes, films, and
  drops.  {\em Soft Matter}  {\bf 2007}, {\em 3}, 1349 -- 1368.

\bibitem[7]{deGennes1995}
de~Gennes, P.; Prost, J. The Physics of Liquid Crystals.second;   Oxford
  University Press: New York, 1995.

\bibitem[8]{Mukherjee2001}
Mukherjee, P.K.; Pleiner, H.; Brand, H.R. Simple Landau model of the
  smectic-A-isotropic phase transition..  {\em European Physical Journal E:
  Soft Matter}  {\bf 2001}, {\em 4}, 293--297.

\bibitem[9]{Abukhdeir2007}
Abukhdeir, N.; Rey, A.  Nonlinear Model for the Isotropic/Smectic A Phase
  Transition. In: {\em Modelling and Simulation},   : , 2007; .

\bibitem[10]{Abukhdeir2008a}
Abukhdeir, N.; Rey, A. Defect kinetics and dynamics of pattern coarsening in a
  two-dimensional smectic-A system.  {\em New Journal of Physics}  {\bf 2008},
  {\em 10} (6), 063025 (17pp).

\bibitem[11]{Abukhdeir2008}
Abukhdeir, N.; Rey, A. Simulation of surface-enhanced ordering in smectic
  films.  {\em Solid State Phenomena}  {\bf 2008}, {\em 139}, 135--140.

\bibitem[12]{Abukhdeir2008c}
Abukhdeir, N.; Rey, A. Simulation of spherulite growth using a comprehensive
  approach to modeling the first-order isotropic/smectic-A mesophase
  transition.  {\em Arxiv preprint arXiv:0807.4525}  {\bf 2008}.  Submitted to
  Communications in Computational Physics July 2008, manuscript ID
  CFluids08-01.

\bibitem[13]{Rey2008}
Rey, A.; Abukhdeir, N. Mechanical Model for Filament Buckling and Growth by
  Phase Ordering.  {\em Langmuir}  {\bf 2008}, {\em 24} (3), 662--665.

\bibitem[14]{Rey2008a}
Rey, A.; Abukhdeir, N. Flow perturbation model for filament buckling.  {\em
  Journal of Non-Newtonian Fluid Mechanics}  {\bf 2008}, {\em 153} (2-3),
  177--182.

\bibitem[15]{Harrison2000}
Harrison, C.; Adamson, D.H.; Cheng, Z.; Sebastian, J.M.; Sethuraman, S.; Huse,
  D.A.; Register, R.A.; et~al. Mechanisms of Ordering in Striped Patterns.
  {\em Science}  {\bf 2000}, {\em 290} (5496), 1558--1560.

\bibitem[16]{Coles1979a}
Coles, H.J.; Strazielle, C. The order-disorder phase transition in liquid
  crystals as a function of molecular structure. I. The alkyl cyanobiphenyls..
  {\em Molecular Crystals and Liquid Crystals}  {\bf 1979}, {\em 55}, 237--50.

\bibitem[17]{Rey2002}
Rey, A.; Denn, M. Dynamical Phenomena in Liquid-Crystalline Materials.  {\em
  Annual Review of Fluid Mechanics}  {\bf 2002}, {\em 34} (1), p233 --.

\bibitem[18]{Yan2002}
Yan, J.; Rey, A.D. Texture formation in carbonaceous mesophase fibers.  {\em
  Phys. Rev. E}  {\bf 2002}, {\em 65} (3) (Feb), 031713.

\bibitem[19]{Toledano1987}
Toledano, J.C.; Toledano, P. The Landau Theory of Phase Transitions:
  Application to Structural, Incommensurate, Magnetic, and Liquid Crystal
  Systems (World Scientific Lecture Notes in Physics).  : , 1987.

\bibitem[20]{Brand2001}
Brand, H.R.; Mukherjee, P.K.; Pleiner, H. Macroscopic dynamics near the
  isotropic-smectic-A phase transition..  {\em Physical Review E: Statistical,
  Nonlinear, and Soft Matter Physics}  {\bf 2001}, {\em 63},
  061708/1--061708/6.

\bibitem[21]{Mukherjee2002a}
Mukherjee, P.K.; Pleiner, H.; Brand, H.R. Landau model of the smectic
  C--isotropic phase transition.  {\em The Journal of Chemical Physics}  {\bf
  2002}, {\em 117} (16), 7788--7792.

\bibitem[22]{Biscari2007}
Biscari, P.; Calderer, M.; Terentjev, E. Landau de Gennes theory of
  isotropic-nematic-smectic liquid crystal transitions.  {\em Phys Rev E Stat
  Nonlin Soft Matter Phys}  {\bf 2007}, {\em 75} (5) (May), 051707.

\bibitem[23]{Barbero2000}
Barbero, G.; Evangelista, L.R. An Elementary Course on the Continuum Theory for
  Nematic Liquid Crystals (Series on Liquid Crystals , Vol 3).  : , 2000.

\bibitem[24]{Ambrozic2004}
Ambrozic, M.; Kralj, S.; Sluckin, T.J.; Zumer, S.; et~al. Annihilation of edge
  dislocations in smectic-A liquid crystals..  {\em Physical Review E:
  Statistical, Nonlinear, and Soft Matter Physics}  {\bf 2004}, {\em 70},
  051704/1--051704/12.

\bibitem[25]{Urban2005}
Urban, S.; Przedmojski, J.; Czub, J. X-ray studies of the layer thickness in
  smectic phases..  {\em Liquid Crystals}  {\bf 2005}, {\em 32}, 619--624.

\bibitem[26]{Wincure2006}
Wincure, B.; Rey, A. Interfacial nematodynamics of heterogeneous curved
  isotropic-nematic moving fronts.  {\em The Journal of Chemical Physics}  {\bf
  2006}, {\em 124} (24), 244902.

\bibitem[27]{Wincure2007}
Wincure, B.; Rey, A. Growth and structure of nematic spherulites under shallow
  thermal quenches.  {\em Continuum Mechanics and Thermodynamics}  {\bf 2007},
  {\em 19} (1) (Jun.), 37--58.

\bibitem[28]{Wincure2007a}
Wincure, B.; Rey, A. Nanoscale Analysis of Defect Shedding from Liquid Crystal
  Interfaces.  {\em Nano Letters}  {\bf 2007}, {\em 7} (6), 1474--1479.

\bibitem[29]{Wincure2007b}
Wincure, B.; Rey, A.D. Computational modelling of nematic phase ordering by
  film and droplet growth over heterogeneous substrates.  {\em Liquid Crystals}
   {\bf 2007}, {\em 34} (12), 1397--1413.

\bibitem[30]{Kirk2006}
Kirk, B.; Peterson, J.W.; Stogner, R.H.; et~al. {\texttt{libMesh}: A C++
  Library for Parallel Adaptive Mesh Refinement/Coarsening Simulations}.  {\em
  Engineering with Computers}  {\bf 2006}, {\em 22} (3--4), 237--254.

\bibitem[31]{Friedel1910}
Friedel, G.; Grandjean, F. {Observation geometriques sur les liquidesa coniques
  focales}.  {\em Bull. Soc. Fr. Mineral}  {\bf 1910}, {\em 33}, 409--465.

\bibitem[32]{Dierking2003}
Dierking, I.; Russell, C. Universal scaling laws for the anisotropic growth of
  SmA liquid crystal batonnets..  {\em Physica B (Amsterdam, Neth.)}  {\bf
  2003}, {\em 325}, 281--286.

\bibitem[33]{Dierking2003a}
Dierking, I. Textures of Liquid Crystals.  : , 2003.

\bibitem[34]{Fournier1991}
Fournier, J.; Durand, G. Focal conic faceting in smectic-A liquid crystals.
  {\em Journal de Physique II}  {\bf 1991}, {\em 1} (Jul.), 845--870.

\bibitem[35]{Virga1994a}
Virga, E. Variational Theories for Liquid Crystals.  : , 1994.

\bibitem[36]{Yokoyama1997}
Yokoyama, H., Interfaces and Thin Films,  {\em Handbook of Liquid Crystal
  Research }; Oxford University Press, 1997; p. 179.

\bibitem[37]{Abukhdeir2008b}
Abukhdeir, N.; Soule?, E.; Rey, A. Non-Isothermal Model for Nematic Spherulite
  Growth.  {\em Langmuir}  {\bf 2008}, {\em 24} (23), 13605--13613.

\bibitem[38]{Chandrasekhar1992}
Chandrasekhar, S. Liquid Crystals.second;   Cambridge University Press:
  Cambridge, 1992.

\bibitem[39]{Ishikawa2001}
Ishikawa, T.; Lavrentovich, O. {\em Defects and undulation in layered liquid
  crystals }; Vol. ~43, 2001; .

\bibitem[40]{Rey2000}
Rey, A. Pitch Contributions to the Cholesteric-Isotropic Interfacial Tension.
  {\em Macromolecules}  {\bf 2000}, {\em 33} (25), 9468--9470.

\bibitem[41]{Prinsen2003}
Prinsen, P.; van~der Schoot, P. Shape and director-field transformation of
  tactoids.  {\em Phys. Rev. E}  {\bf 2003}, {\em 68} (2) (Aug), 021701.

\bibitem[42]{Prinsen2004}
Prinsen, P.; Schoot, P. Continuous director-field transformation of nematic
  tactoids.  {\em The European Physical Journal E: Soft Matter and Biological
  Physics}  {\bf 2004}, {\em 13} (1) (Jan.), 35--41.

\bibitem[43]{Storti2008}
Storti, M.; Nigro, N.; Paz, R., PETSc-FEM A general purpose, parallel,
  multi-physics FEM program, http://www.cimec.org.ar/petscfem/  International
  Center of Computational Methods in Engineering (CIMEC), 2008.

\end{thebibliography}

\label{lastpage}

\end{document}